\documentclass[12pt]{article}

\PassOptionsToPackage{numbers, compress}{natbib}

\usepackage[preprint]{neurips_2019}

\usepackage[utf8]{inputenc} 
\usepackage[T1]{fontenc}    
\usepackage{hyperref}       
\usepackage{url}            
\usepackage{booktabs}       
\usepackage{amsfonts}       
\usepackage{nicefrac}       
\usepackage{microtype}      
\usepackage{caption}
\usepackage{subcaption}
\usepackage{graphicx}
\usepackage{amsmath,amssymb,amsthm}
\usepackage{dsfont}
\usepackage{bm}
\usepackage{geometry}
\usepackage{afterpage}
\usepackage{algorithm}
\usepackage{multirow}
\usepackage{algorithmic}

\title{Application of Time Series Analysis to Traffic Accidents in Los Angeles}

\author{
  Qinghao Ye$^{1}$\thanks{Denotes corresponding author.} , Kaiyuan Hu$^1$, \textbf{Yizhe Wang}$^2$\\
 $^1$ School of Computer Science and Technology, Hangzhou Dianzi University, China\\
 $^2$ Department of Statistic, University of California Riverside, CA, USA \\
  \texttt{\{16205234, 16051514\}@hdu.edu.cn}\quad \texttt{ywang668@ucr.edu} \\
 }

\begin{document}

\maketitle

\begin{abstract}
  With the improvements of Los Angeles in many aspects, people in mounting numbers tend to live or travel to the city. The primary objective of this paper is to apply a set of methods for the time series analysis of traffic accidents in Los Angeles in the past few years. The number of traffic accidents, collected from 2010 to 2019 monthly reveals that the traffic accident happens seasonally and increasing with fluctuation. This paper utilizes the ensemble methods to combine several different methods to model the data from various perspectives, which can lead to better forecasting accuracy. The IMA(1, 1), ETS(A, N, A), and two models with Fourier items are failed in independence assumption checking. However, the Online Gradient Descent (OGD) model generated by the ensemble method shows the perfect fit in the data modeling, which is the state-of-the-art model among our candidate models. Therefore, it can be easier to accurately forecast future traffic accidents based on previous data through our model, which can help designers to make better plans. 
\end{abstract}

\section{Introduction}
As the country with the highest car ownership and the second largest car ownership per capita in the world, the traffic problem in the United States has always been the focus of people. According to The World Bank's statistics "World Bank Data: Motor vehicles (per 1,000 people)" \cite{motor}, the number of cars per thousand people in the United States reached 910 in 2017, and the total amount reached 255,009,283 \cite{capitoltires}.

Los Angeles is the largest city in California and the second largest city in the United States, second only to New York City. The city covers an area of 469.1 square miles (1214.9 square kilometers). The Los Angeles metropolitan area consists of Los Angeles, Long Beach and Anaheim, has a population of approximately 13.31 million, and the greater Los Angeles area contains 18.7 million residents totally \cite{us2016annual}. Both statistical data are roughly equal to New York.

Today, Los Angeles has developed into one of the greatest international metropolis, with highly developed industry of culture, science, technology, sports, international trade and education. Among all these industries, the thriving development of the culture and entertainment industry, which produces great films, TV series, and music, plays the most significant role, also forms the foundation of Los Angeles' international reputation and global status.
	
For entertainment industry, Hollywood is the entertainment center of Los Angeles. It takes about 40 minutes by bus from downtown to Hollywood. Since, 1911, the very year the first film company established in Hollywood, the area has become the world's film center. With so many stars and great movies, it is safe to say that Hollywood is the pioneer of global film industry. Celebrity Avenue is a world famous sight located in Hollywood, records more than 2000 highly-achieved celebrities in the fields of film, broadcasting and music.

When it comes to the education industry, the education in Los Angeles is well-known all over the world. Los Angeles is home to the world-famous California Institute of Technology, the University of California Los Angeles and University of Southern California. Apart from famous universities, Los Angeles Public Library, Huntington Library and Getty Museum are also important cultural and educational facilities. 

In the sports industry, Los Angeles has also developed a very great environment for sports, provides opportunities for all kinds of sports. Los Angeles Angels, a Major League Baseball team in Los Angeles, California, part of the Pacific Coast League. Los Angeles Lakers, a basketball club founded in Minneapolis in 1947 but developed in Los Angeles, won sixteen championships in National Basketball Association. Also, Los Angeles held Olympic Summer Games twice in 1932 and 1984.

Due to the high level of development in various industries, Los Angeles, as an international metropolis, inevitably attracts a large number of immigrants to settle down. Among all the residents in Los Angeles, 11.2\% of them are immigrants from other countries, of which Chinese are the most, accounting for 7.1\%.

As the population continues to expand, the traffic, inevitably become a serious question. Simultaneously, Los Angeles government is constantly improving the urban transportation system: Los Angeles has one of the world's largest highway systems, yet it is still the top 10 worst traffic city in the United States \cite{sorensen2009moving}. The main roads connecting Los Angeles and surrounding cities include Interstate 5, Interstate 10, and US National Highway 101. The Los Angeles County Metropolitan Transportation Authority and other agencies operate numerous bus lines as well as subway and light rail lines. Los Angeles rail transit includes red and purple subway lines, as well as gold, blue, expo and green light rail lines. The orange and silver lines are BRT lines with stops and frequencies similar to light rails. The main train station in Los Angeles is the Union Station in the northeast of the city center, the center of the Southern California Metropolis Railway. The system connects Los Angeles with all neighboring counties and many suburbs. 

Los Angeles metropolis (surrounding areas included) is the second largest metropolitan area in the United States. According to survey, approximately a third of the total population resides in the City of Los Angeles \cite{ong2005spatial}. However, the annual flow of passengers in the Los Angeles subway is only about 48.7 million. In contrast, in Hong Kong, where the population is much smaller than Los Angeles, the subway transports 1.563 billion passengers a year, almost the 30 times of that of Los Angeles.

A large number of population will inevitably bring a lot of traffic demand, but the low public transportation utilization rate has caused Los Angeles to become one of the ten most congested cities in the United States. What followed is the inevitable traffic accidents and safety problems. Our group selected traffic collision reported to LAPD as our data set, hoping to have a better understanding of the traffic condition in LA.

In short, the main contribution of this work can be highlighted in following aspects:
\begin{itemize}
    \item By adding the Fourier items, the complex seasonality can be well modeled by an intuitive manner. The proposed ensembled model outperforms other methods by a large margin.
    \item Model diagnostics are fully conducted, and the components of ensemble model are also investigated.
\end{itemize}{}

The remaining parts are organized as follows. Section \ref{sec:dataset} generally explores the time series we are interested in and Section \ref{sec:methods} introduces plenty of time series modeling methods. Then, Section \ref{sec:experiments} describes a number of experiments carried out on the training and validation set and reports the models' performance. After that, Section \ref{sec:diagnostic} uses several diagnostic techniques to evaluate the validity of the candidate models. Finally, we conduct the forecast in Section \ref{sec:forecast} and conclude this paper.


\section{Data Exploration}
\label{sec:dataset}
\subsection{Data Description}

\begin{figure}[t!]
     \centering
     \begin{subfigure}[t]{.5\textwidth}
        \centering
        \includegraphics[width=0.95\textwidth]{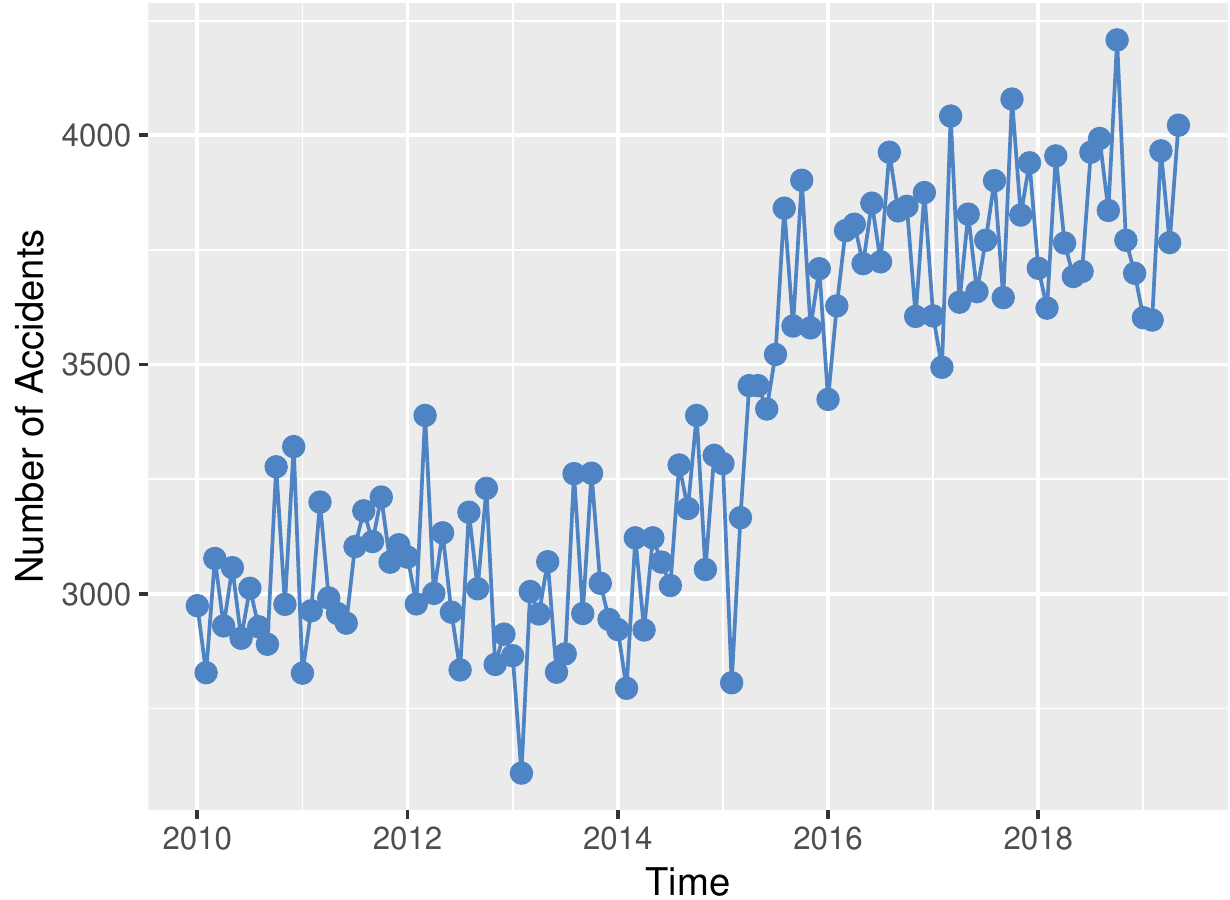}
        \caption{Original series.}
    \end{subfigure}%
    ~
    \begin{subfigure}[t]{.45\textwidth}
        \centering
        \includegraphics[width=0.95\textwidth]{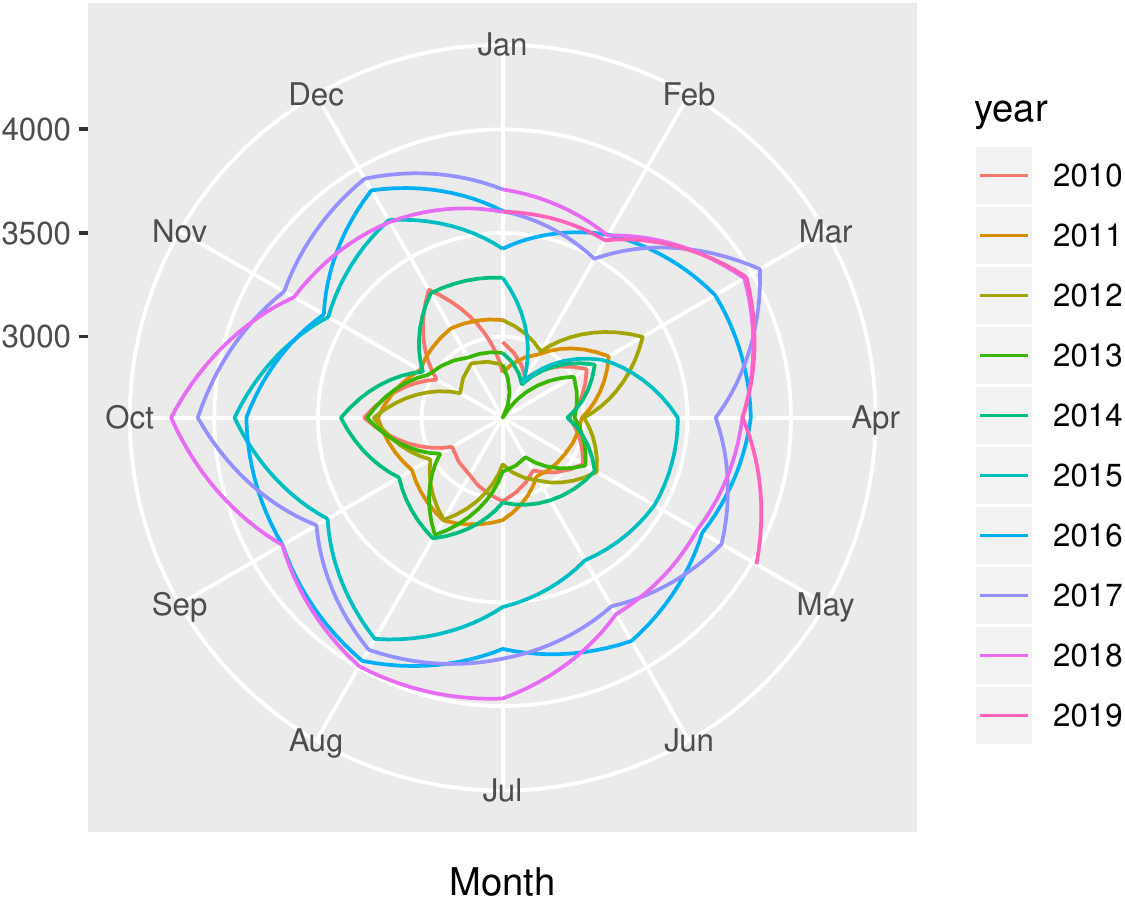}
        \caption{Seasonal plot of original series distribution.}
    \end{subfigure}
     \caption{Time series of traffic accidents in Los Angeles.}
    \label{fig:data}
\end{figure}

The Los Angeles traffic accidents are collected by City of Los Angeles from 2010 to 2019 monthly. The corresponding values are integer. The data set contains 113 data points. The plot of the data and data distribution are demonstrated in Figure \ref{fig:data}. It shows that accidents are more likely to happen in October and less likely in February.

\subsection{Box-Cox Transformation}
\begin{figure}[t!]
     \centering
     \begin{subfigure}[t]{.5\textwidth}
        \centering
        \includegraphics[width=0.95\textwidth]{./Original_series.pdf}
        \caption{Origin series without Box-Cox transformation.}
    \end{subfigure}%
    ~
    \begin{subfigure}[t]{.5\textwidth}
        \centering
        \includegraphics[width=0.95\textwidth]{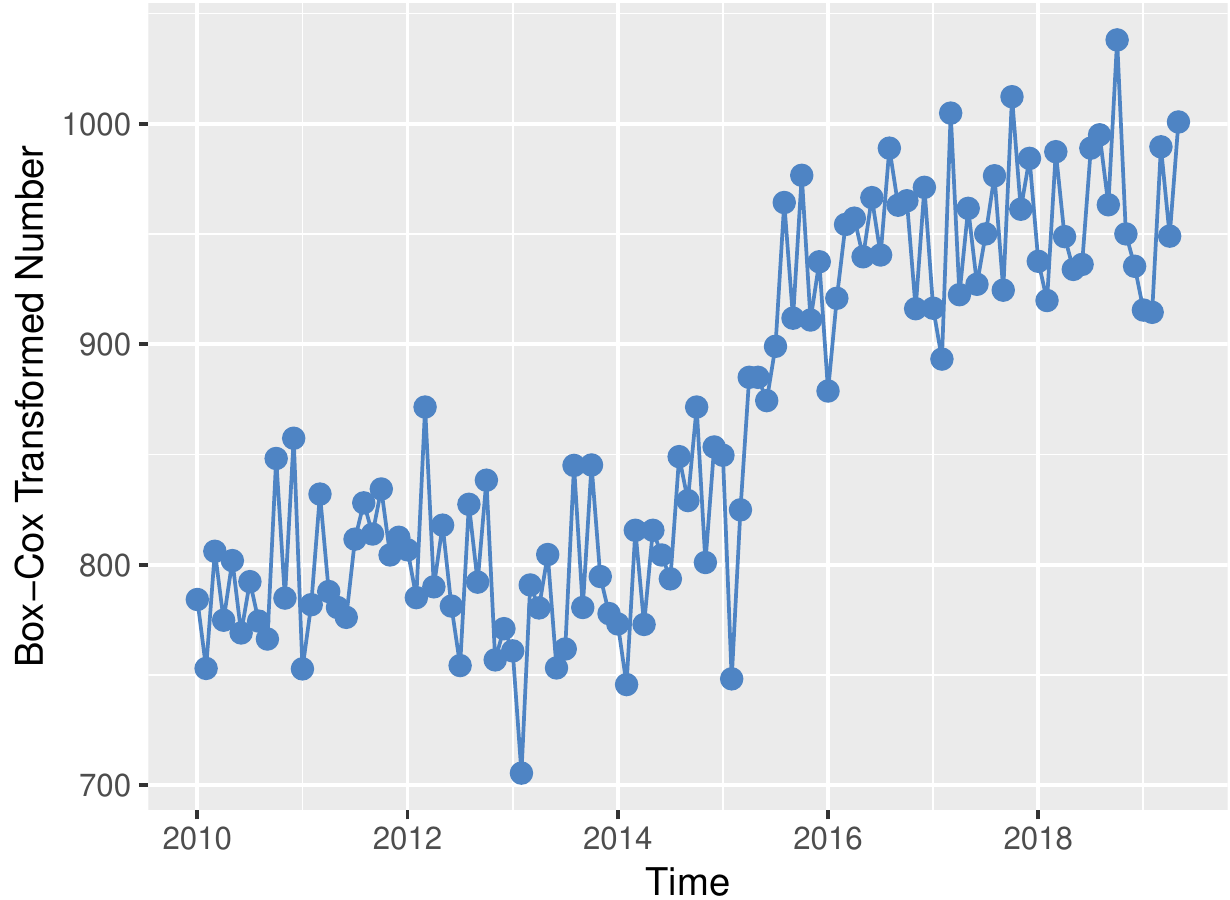}
        \caption{Box-Cox transformed series with $\lambda = 0.60377$.}
    \end{subfigure}
     \caption{Number of Traffic Accidents in Los Angeles data, which is a daily time series. The original series and transformed series are shown in the above figures.}
    \label{fig:data_coxbox}
\end{figure}

In the real-world scenario, the distribution of running times data may not be normally distributed, which results in the distribution being asymmetrical leading to a bias in the model. To deal with the problem, George Box et al. \cite{box1964analysis} propose a data transformation technique named Box-Cox Transformation which helps to stabilize the variance of the series. The mathematical expression is written as
\begin{align}
    y_t(\lambda) = \begin{cases}
    \frac{y_t^\lambda - 1}{\lambda} \quad \quad & \text{if }\lambda \neq 0, \\
    \log y_t \quad \quad & \text{if }\lambda = 0,
    \end{cases}
\end{align}{}
where $\lambda$ is the power coefficient of the transformation that we need to select. We can find out that when $\lambda = 1$, the whole series is identical only with shift. When $\lambda = 0$, the transformation is the logarithm. However, it is difficult to choose the parameter $\lambda$ since the searching space is infinite. Therefore, Guerrero \citep{guerrero1993time} proposes a method to estimate the value of $\lambda$ which aims to minimize the coefficient of variation for subsequences of the original series. Figure \ref{fig:data_coxbox} shows the original series and transformed series with $\lambda = 0.60377$.

When we tend to inverse the process of Box-Cox transformation, the transformed predicted value will not be the mean of the forecast distribution but the median. Since the median cannot be added up, so we should use bias adjustments to find the corrected version of the inverse of Box-Cox transformation, which is formulated as follows
\begin{align}
y_t =
  \begin{cases}
     \exp(y_t(\lambda))\left[1 + \frac{\sigma_h^2}{2}\right] & \text{if $\lambda=0$;}\\
     (\lambda y_t(\lambda)+1)^{1/\lambda}\left[1 + \frac{\sigma_h^2(1-\lambda)}{2(\lambda y_t(\lambda)+1)^{2}}\right] & \text{otherwise;}
  \end{cases}
\end{align}
where $\sigma_h^2$ is the $h$-step predict variance, the larger variance, the larger difference between mean and median.

\subsection{Data Characteristics}
Understanding of time series data is an essential part of the time series analysis process. Before we apply the model to fit the data, we should explore the statistical characteristics of the time series data which contains seasonality, trend, noisy, chaos, etc. In this part, we consider six main statistical characteristics of the given data. In order to vividly understand the data, we normalize the different metrics into $[0,1]$ to the significance of the characteristics that are presented or not. The larger the normalized value, the more significance it shows.

\paragraph{Trend and Seasonality}
Because decomposition of time series can be applied to measure the trend and seasonality in time series. Following the STL decomposition \cite{cleveland1990stl}, we decompose the series as $y_t = T_t + S_t + R_t$, which are trend component, seasonal component, and remainder component, respectively. Here, we utilize the strength of trend $F_T$ and the strength of seasonality $F_S$ proposed by Wang et al. \cite{wang2006characteristic} to measure the trend and seasonality. The mathematical expressions are presented as 
\begin{equation*}
    F_T = \max \left(0, 1-\frac{Var(R_t)}{Var(T_t + R_t)}\right), \quad F_S = \max \left(0, 1-\frac{Var(R_t)}{Var(S_t + R_t)}\right).  
\end{equation*}
For strongly trend or seasonal series, there should be have more variation in trend or seasonality than remainder component leading to the larger $F_T$ or $F_S$.

\paragraph{Serial Correlation}
To estimate the serial correlation, we use the Box-Pierce statistics $Q_h$ in our time series analysis process. Box-Pierce statistic is formulated as $Q_h = n \sum_{k=1}^h \gamma_k^2$, where $\gamma_k$ is the autocorrelation with time lag $k$, $n$ is the length of the sequence, and $h$ is the maximum lag we want to measure (we set $h = 20$).

\paragraph{Skewness}
When analyzing the data, we are interested in the symmetry of the distribution of the data. To quantify the the degree of asymmetry of sequence values around the mean value, the skewness coefficient $S = \frac{1}{n\sigma^3}\sum_{t=1}^n |y_t - \bar{y}_t|^3$ adopted. It should be noticed that symmetric data lead to the skewness coefficient $S$ close to zero. Oppositely, if $S$ is large, the data would more likely be asymmetrical around the mean value. 

\paragraph{Kurtosis}
In order to measure whether the time series is peaked or flat, we use Kurtosis coefficient $K = \frac{1}{n\sigma^4}\sum_{t=1}^n(y_t - \bar{y}_t)^4$. For low kurtosis coefficient $K$, the series tend to become flat rather than a sharp peak near the mean. So we can use the Kurtosis coefficient to measure whether it has a heavy tail or not.

\paragraph{Self-similarity}
Since the traditional autoregressive models are only capable of capturing short-range correlation of the time series data \cite{hosking1984modeling}. To measure whether the data is self-similarity (long-range dependent), we use self-similarity parameter Hurst exponent $H$ \cite{terasvirta1996power} to quantify the long-range dependence. We follow Wang et al. \cite{wang2006characteristic} to achieve the value. The higher the value $H$, the more long-range dependence the data exhibits.

\begin{table}[]
\centering
\begin{tabular}{c|c|c|c|c|c}
\hline\hline
Trend  & Seasonality & Serial Correlation & Skewness & Kurtosis & Self-similarity \\ \hline
0.9370 & 0.6215      & 0.9852             & 0.0567   & 0.0039   & 0.9818          \\ \hline
\end{tabular}
\vspace{0.05in}
\caption{Scaled statistical characteristics of the traffic accident time series. The higher statistics means higher possibility that the phenomenon would occur.}
\label{tab:test_statistic}
\end{table}

By using these statistical characteristics, we compute these statistics on the time series of traffic accidents in Log Angeles. Then we scale these values into $[0,1]$ for easier interpretation. The scaled statistical characteristics are recorded in Table \ref{tab:test_statistic}. As the table shows, the traffic accident sequence has a trend and slightly seasonality, and we can observe that the series is highly correlated. Additionally, the distribution of the series value is balance since the skewness and Kurtosis coefficient are relatively low, showing the normality assumption of the data. On the other hand, the self-similarity parameter is high showing that the current value in the sequence is correlated with the data from a long time ago. Therefore, we might use some models with the ability to capture long-range dependence.

Besides, we examine the time series of traffic accidents is stationary or not. We adopt Augmented Dickey-Fuller test (ADF) to verify the result. ADF tests the null hypothesis that there is a unit root is present in the time series. If the time series has a unit root, it means that the testing series is not stationary and has some time-related structure. If the series is not stationary, we can use take the difference to the series several times to make it stationary. The results of ADF test are described in Table \ref{tab:adf}. In the table, we can observe that the p-value is large when we testing original series indicating that we failed to reject null hypothesis, which means that the original series has a unit root and is non-stationary. On the other hand, after taking the first-order difference of the series, it results in the small p-value showing the differenced series does not have a unit root and is stationary, which suggests that we should take first-order difference to the series before we applying the stationary model.

\begin{table}[]
\centering
\begin{tabular}{l|c|c|c|c}
\hline\hline
\textbf{Testing Series} & \textbf{Dickey-Fuller} & \textbf{Lag order} & \textbf{p-value} & \textbf{Stationary} \\ \hline
$\{y_t\}_{t=1}^n$             & -1.6938                & 4                  & 0.7022           & No                  \\
$\{y_t - y_{t-1}\}_{t=2}^{n}$   & -5.0588                & 4                  & < 0.01       & Yes     \\ \hline
\end{tabular}
\vspace{0.05in}
\caption{Augmented Dickey-Fuller test results for the traffic accidents series.}
\label{tab:adf}
\end{table}

\begin{figure}[t!]
     \centering
     \begin{subfigure}[t]{0.99\textwidth}
        \centering
        \includegraphics[width=0.95\textwidth]{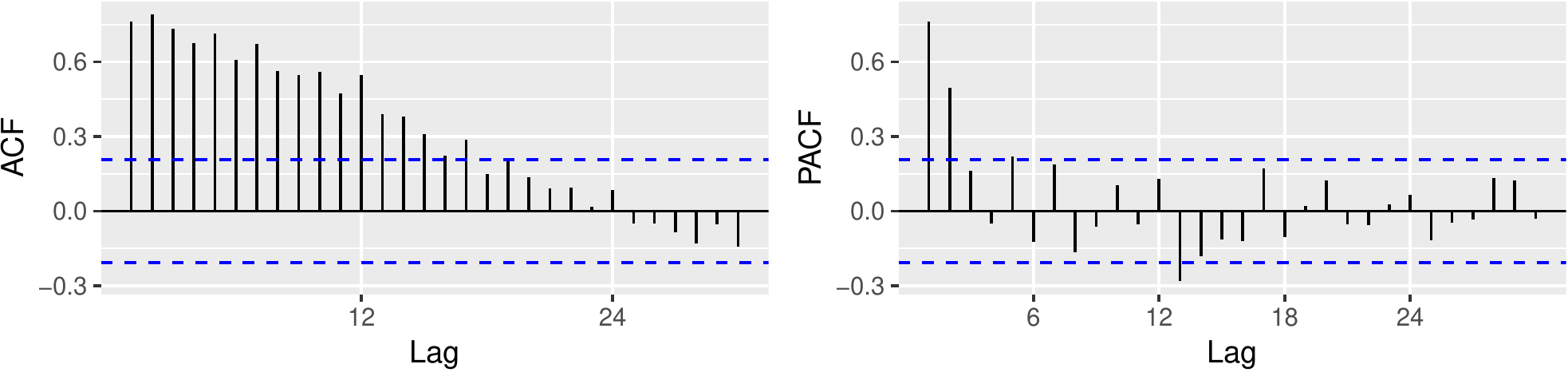}
        \caption{ACF plot and PACF plot for the original series.}
        \label{fig:acf_original}
    \end{subfigure}%
    \vspace{0.05in}
    \begin{subfigure}[t]{0.99\textwidth}
        \centering
        \includegraphics[width=0.95\textwidth]{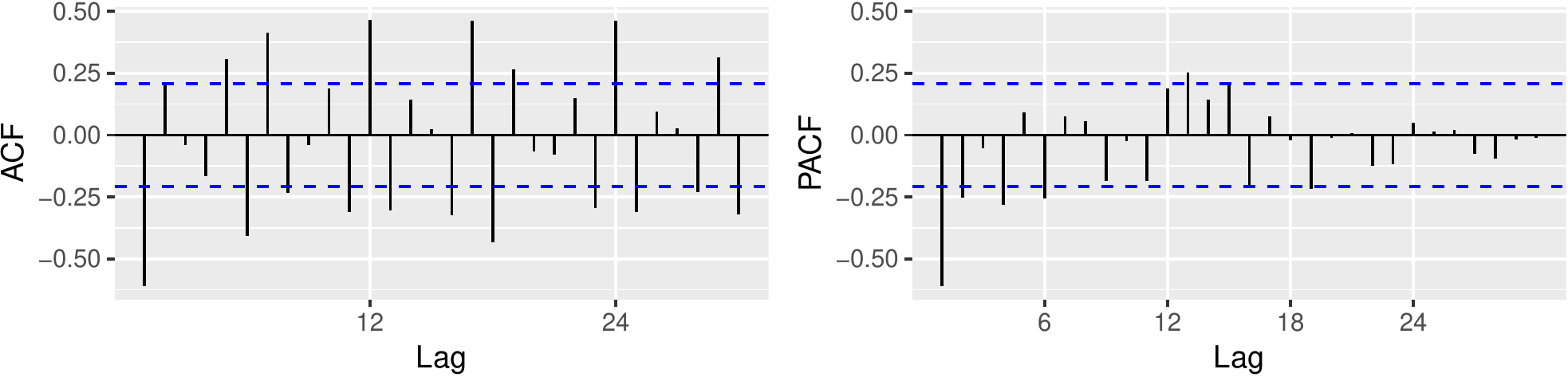}
        \caption{ACF plot and PACF plot for the first-order differenced series.}
        \label{fig:acf_diff}
    \end{subfigure}
     \caption{ACF plots and PACF plots for original and first-order differented series, which are used for choosing the proper AR or MA family model.}
    \label{fig:acf_pacf}
\end{figure}

In addition, we compute the autocorrelation and partial autocorrelation among original and first-order differenced series, which is presented in Figure \ref{fig:acf_pacf}. In Figure \ref{fig:acf_original}, we can clearly see from the ACF plot that the time series is strongly correlated with its past, showing that it is not proper to use ARMA family model. Therefore, after taking the first-order difference, we still can clearly see the data is highly correlated since there are a large quantity of spikes in the ACF plot. But according to the result of ADF test above, we know the first-order differenced series is stationary, which means that we can fit ARMA family model to the data. From the PACF plot in Figure \ref{fig:acf_diff}, we empirically regard its shape as the MA(1) process due to the fact that only lag 1 shows negative strong partial autocorrelation. Therefore, when we fit MA model to the first-order differenced series, we may choose $q = 1$ as the order.

When it comes to ARMA model choosing, we also calculate the extended autocorrelation function (EACF) of first-order differenced series in order to select the proper order $(p, q)$. The calculated extended autocorrelations are demonstrated in Table \ref{tab:eacf}. From the table, we can observe plenty of false negative results. However, it can be clearly seen that the triangle of "o"s starts at $(p,q) = (1,1)$ meaning that we might use ARIMA(1,1,1) model.

\begin{table}[]
\centering
\begin{tabular}{ccccccccccccccc}
\hline\hline
\multicolumn{15}{c}{MA}                                        \\ \cline{2-15} 
AR & 0 & 1 & 2 & 3 & 4 & 5 & 6 & 7 & 8 & 9 & 10 & 11 & 12 & 13 \\ \hline
0  & x & o & o & o & x & x & x & x & o & o & x  & x  & x  & o  \\
1  & x & o & o & o & o & o & x & x & o & o & o  & x  & o  & o  \\
2  & x & o & x & o & o & o & o & o & o & o & o  & x  & o  & o  \\
3  & o & x & o & o & o & o & o & o & o & o & o  & x  & o  & o  \\
4  & x & x & x & o & x & o & o & o & o & o & o  & o  & o  & o  \\
5  & x & x & x & o & o & o & o & o & o & o & o  & o  & o  & o  \\
6  & x & x & x & o & o & o & o & o & o & o & o  & o  & o  & o  \\
7  & x & o & x & o & x & o & x & o & o & o & o  & o  & o  & o  \\ \hline
\end{tabular}
\vspace{0.05in}
\caption{EACF table for the first-order differenced series used for choosing the proper orders $(p, q)$ for ARMA model.}
\label{tab:eacf}
\end{table}

\section{The Proposed Approaches}
\label{sec:methods}
In this section, we first introduce a variety of parametric methods to model the time series data, which employ parameters to model the linear relation within residuals or previous data, like AutoRegressive model (AR), Moving Average model (MA), and AutoRegressive Integrated Moving Average model (ARIMA). After introducing ARIMA family model, we utilize bootstrapping sampling methods to augment the data set and perform the exponential smoothing methods to model the series. Beyond these two family models, we also perform Fourier terms to enhance the models' ability to model the complex and multiple seasonality. At last, we propose the final model by adapting ensemble learning technique from the statistical learning area.

\subsection{AutoRegressive Model (AR)}
Intuitively, the sequence data tends to show some correlations with the previous data or trend, which means that future trend might base on the previous data points. Therefore, we can forecast the variable of interest using a linear combination of past values of the variable. Yule \cite{udny1927method} formulated an autoregressive model with the order $p$ (i.e. AR($p$)) as
\begin{equation}
    y_t = c + \sum_{i=1}^p \psi_i y_{t-i} + \epsilon_t,
\end{equation}{}
where $\epsilon_t$ is the white noise at time step $t$, and $c$ is the drift value based on previous observed data points.

Once we have the observed data, we can estimate the parameters $\{\psi_i\}_{i = 1}^p$ using the ordinary least squares which can be derived by the following the equation
\begin{equation}
    \mathop{\arg \min}_{\psi_i, \cdots, \psi_p} \sum_{j=1}^n (\hat{y}_j - y_j)^2.
\label{eq:ar_objective}
\end{equation}{}
The Eq. (\ref{eq:ar_objective}) can be solved by the Yule-Walker equations \cite{udny1927method}. 

\subsection{Moving Average (MA)}
Instead of using the past values from the sequence, Slutzky \cite{slutzky1937summation} utilized past forecast errors $\epsilon_t$ to model the sequence. The proposed moving average model of order $q$ (i.e. MA($q$)) is represented as  
\begin{equation}
    y_t = c + \sum_{i=1}^p \theta_i \epsilon_{t-i} + \epsilon_t,
\end{equation}{}
where $\epsilon_t$ is the white noise at time step $t$, and $c$ is the drift value for the sequence.

Besides, it is trivial to verify that AR($p$) model can be formulated as MA($\infty$), since
\begin{equation}
    y_t = \eta y_{t-1} + \epsilon_t = \eta^2 y_{t-2} + \eta \epsilon_{t-1} + \epsilon_t = \cdots = \sum_{i = 0}^\infty \eta^i \epsilon_{t-i}
\end{equation}{}

\subsection{AutoRegressive Integrated Moving Average (ARIMA)}
By combining AR model and MA model, Makridakis et al. proposed a non-seasonal autoregressive integrated moving average model (i.e. ARIMA($p,d,q$)) which is written as
\begin{equation}
    y_t' = c + \sum_{i=1}^p \psi_i y_{t-i}' + \sum_{j = 1}^q \theta_j\epsilon_{t-j} + \epsilon_t,
\end{equation}{}
where $\{y_t'\}_{t = 1}^{n-d}$ denotes the $d$ times differenced series with respected to series $\{y_t\}_{t = 1}^{n}$.

Given the $p, d, q$, we can estimate the parameters $\Theta = \{c, \phi_1, \cdots, \phi_p, \theta_1, \cdots, \theta_q\}$ in the ARIMA($p,d,q$) model by maximizing the likelihood of the probability of obtaining the observed data, which is equivalent to minimizing the summation of forecast errors as 
\begin{equation}
    \min_{\Theta} \sum_{i = 1}^n \epsilon_t^2.
\end{equation}{}

Furthermore, in order to choose the proper order of $p$ and $q$ for ARIMA model, we use the automatic model selection technique proposed by Bozdogan Hamparsum \cite{bozdogan1987model}, which adapts Akaike’s Information Criterion (AIC) \cite{aho2014model}, corrected AIC (AICc), and Bayesian Information Criterion (BIC) \cite{aho2014model} as the criterion to select the proper order $p$ and $q$. The AIC, AICc, and BIC are formulated as
\begin{align}
    & AIC = -2\log (\hat{L}) + 2(p+q+\mathds{1}\{c\neq0\} + 1), \\
    & AICc = AIC + \frac{2(p+q+\mathds{1}\{c\neq0\}+1)(p+q+\mathds{1}\{c\neq0\}+2)}{n-p-q-\mathds{1}\{c\neq0\}-2},\\
    & BIC = AIC + (\log(n) - 2)(p + q + \mathds{1}\{c\neq0\} + 1),
\end{align}{}
where $\mathds{1}\{c\neq0\} = 1$ when the condition $c\neq 0$ holds true, else $\mathds{1}\{c \neq 0\} = 0$.  $\hat{L}$ is the likelihood of the sequence. According to the criteria of model selection \cite{bozdogan1987model}, a good model should have low values of AIC, AICc, and BIC. We follow this criteria to select the most proper order $p$ and $q$. Hyndman et al. \cite{hyndman2007automatic} propose an algorithm to decide the proper orders, which is summarized in Algorithm \ref{alg:arima}.

\begin{algorithm}
\caption{Hyndman-Khandakar Algorithm}\label{alg:arima}
\begin{algorithmic}[1]

\REQUIRE ~~\\
Training Sequence: $\{y_t\}_{t = 1}^{n}$.

\ENSURE ~~\\
Optimal Model Configuration: $\mathcal{P} = \{ p^*, d^*, q^*\}$. \\

\STATE Using KPSS Test \cite{kwiatkowski1992testing} to determine differences time $0\leq d^*\leq 2 $.
\STATE Differencing the $\{y_t\}_{t = 1}^{n}$ $d$ times to get the differenced data $\{y_t'\}_{t = 1}^{n-d^*}$.
\STATE $\{{AICc}^*, p, q\} \leftarrow \{+\infty, 0, 0\}$.
\STATE Initialize base model ARIMA($p,d^*,q$).
\REPEAT
    \IF{$AICc_{ARIMA(p,d^*,q)} \leq {AICc}^*$}
        \STATE $\mathcal{P} \leftarrow \{p, d^*, q\}$.
        \STATE ${AICc}^* \leftarrow AICc_{ARIMA(p,d^*,q)}$.
    \ENDIF
    \STATE Randomly adding or minusing $p$ or $q$ with $1$ ($p, q \geq 0$) by grid search method.
\UNTIL{No lower AICc can be found.}
\RETURN $\mathcal{P}$.

\end{algorithmic}
\end{algorithm}

Simultaneously, we can incorporate the seasonality into the non-seasonal ARIMA to adapt the time series with seasonality.


\subsubsection{Dynamic Harmonic Regression}
Sometimes the series with long seasonal periods or complex seasonal patterns, the ARIMA model and Seasonal ARIMA model cannot capture the periodical pattern well. Meanwhile, if the series has more than one seasonal periods, it would be difficult to model the patterns only using ARIMA model. Therefore, to handle this problem, we extend ARIMA model with dynamic harmonic regression (DHR) due to the periodic seasonality can be handled using pairs of Fourier terms. We can formulate this model as
\begin{align}
    y_t &= \beta_0 + \sum_{k=1}^K [\alpha_ks_k(t) + \gamma_k c_k(t)] + X_t, \\
    s_k(t) &= \sin \left(\frac{2\pi kt}{m}\right),\\
    c_k(t) &= \cos \left(\frac{2\pi kt}{m}\right),
\end{align}{}
where $\beta_0$ is the intercept term,  $\{\alpha_k\}_{k=1}^K, \{\gamma_k\}_{k=1}^K$ are estimated regression coefficients, and $m$ is seasonal period. $X_t$ is value of non-seasonal ARIMA($p,d,q$) process at time step $t$. By using Fourier terms, every period function can be approximated by the combination of sine and cosine function for large enough $K$.

\subsection{Exponential Smoothing}
\subsubsection{Simple Exponential Smoothing} Simple exponential smoothing is suitable for the scenario that the sequence shows no clear trend or seasonality. Mathematically, it can be written as
\begin{equation}
    \hat{y}_{t+1|t} = \alpha y_t + (1-\alpha) \hat{y}_{t|t-1} = \cdots = \sum_{i = 0}^{n-1}\alpha (1-\alpha)^i y_{t-i} + (1-\alpha)^n \ell_0,
\label{eq:ses}
\end{equation}
where $\ell_0$ is the fitted value at time step $1$ that needs to be estimated, and $\alpha$ is the dampening value with $0\leq \alpha \leq 1$. For the parameter estimation, we following the same protocol for solving parameters in AR model by minimizing Eq. (\ref{eq:ar_objective}). For solving this optimization problem, we can use some iterative optimization approaches such as gradient descent and Newton's method. Moreover, we can divide Eq. (\ref{eq:ses}) into two separate parts as $\hat{y}_{t+1|t} = \ell_t$, and $\ell_t = \alpha y_t + (1-\alpha)\ell_{t-1}$. Here, we can interpret $\ell_t$ as the smoothed value (level) of the sequence at time step $t$.

\subsubsection{Error, Trend, Seasonality (ETS)}
Adapting the idea of simple exponential smoothing, we can model the different aspects of the given series. For example, we are able to model the the trend, seasonal, and reminder components. These components can model long-term direction, periodicity, and error terms. Specially, the trend is usually expressed as the combination of a level term and slope term. Take Holt-Winters model \cite{holt2004forecasting,winters1960forecasting} for instance, this model has addictive trend and additive seasonality, which is formulated by
\begin{align}
    \ell_t &= \alpha(y_t - s_{t-m}) + (1-\alpha)(\ell_{t-1} + b_{t-1}) \\
    b_t &= \beta^* (\ell_t - \ell_{t-1}) + (1 - \beta^*)b_{t-1} \\
    s_t &= \gamma(y_t - \ell_{t-1} - b_{t-1}) + (1-\gamma)s_{t-m}\\
    \hat{y}_{t+h|t} &= \ell_t + hb_t + s_{t-m + [(h-1)\text{ mod }m] + 1} 
\end{align}
where $b_t$ is the slope at time step $t$, $s_t$ denotes the estimate of seasonal component of the sequence at time step $t$, and $m$ is the season number within a year. $\alpha, \beta^*$, and $\gamma$ are the smoothing parameters with $0\leq \alpha, \beta^*$, $\gamma \leq 1$. $h$ is the forecast step.

Moreover, we tend to regard Error, Trend, Seasonality (ETS) as the family of diverse combination of Error, Trend, and Seasonality. To specify, the error can be additive ($A$) or multiplicative ($M$); the trend can be non-existent ($N$), additive, damped additive ($A_d$), multiplicative, or damped multiplicative ($M_d$); the seasonality can be non-existent, additive, or multiplicative. For each combination of three components, we refer to the model as ETS($\cdot$, $\cdot$, $\cdot$). To summarize, there are total 30 different kind of models refer to Hyndman et al. \cite{hyndman2002state}. 

For components selection of ETS family models, we cab adopt the same protocol of ARIMA selection in which utilize the AICc as the criteria. Beside, the initial conditions and smoothing parameters can be estimated by maximizing the likelihood with respect to given data with simplex optimizer \cite{nelder1965simplex}.


\subsubsection{Bootstrapping with ETS}
In this part, we augment our data set with bootstrapping technique. For any time series data, we can use STL decomposition \cite{cleveland1990stl} for seasonal series or loess \cite{cleveland2017local} for non-seasonal series to decompose the series into several components (i.e. trend, remainders, and seasonality only with seasonal series). Because the sequences generated by the STL decomposition are independent and other two terms are changed over time, we can independently simulate the remainder term in STL using bootstrapping method. 

\begin{figure}
     \centering
     \includegraphics[width=0.85\textwidth]{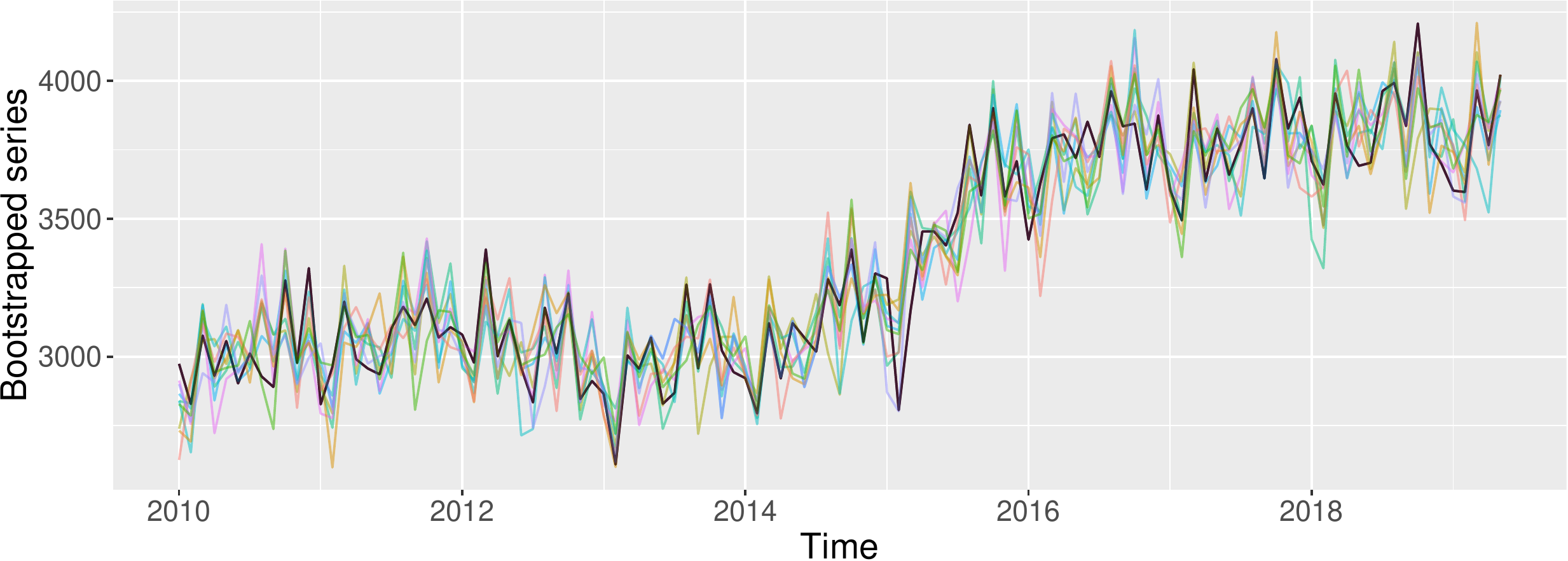}
     \caption{Number of Traffic Accidents in Los Angeles data after the bootstrapping procedure described in Algorithm \ref{alg:bootstrap}. The black line is the original series, and the rest color ones are generated series. The graph is best viewed in color.}
    \label{fig:bootstrapped}
\end{figure}

Therefore, to ensure the stationary property of the series, we bootstrap the remainder of the series after performing STL or loess decomposition. Kunsch proposed MBB process \cite{kunsch1989jackknife} which only requires stationary of the data, showing the desired data length is satisfied by sampling data block with equal size from the series repeatedly. Especially, given a series with length $n$ and the fixed block size $l$, there exists $n-l+1$ blocks at the most. After bootstrapping the remainder, we combine them with the trend and seasonality term generated by STL. Then, the inverse of Box-Cox transformation with bias adjustment is applied to get the final bootstrapped series. The whole process is summarized in Algorithm \ref{alg:bootstrap}. Figure \ref{fig:bootstrapped} shows the series generated by bootstrapping. Empirically, we set the block size $l = 24$ for monthly data since we should the seasonality is observed.

\begin{algorithm}
\caption{Generating Bootstrapped Time Series}\label{alg:bootstrap}
\begin{algorithmic}[1]

\REQUIRE ~~\\
Training Sequence: ts \\
Number of Bootstrapped Times: $Bs$ \\
Block Size: $l$

\ENSURE ~~\\
Bootstrapped Series: ts.boot \\

\STATE $\lambda \leftarrow$ \textbf{BoxCox.lambda}(ts)
\STATE ts.transformed $\leftarrow $ \textbf{BoxCox}(ts, $\lambda$)
\IF{ts is seasonal series}
    \STATE [trend, seasonal, remainder] $\leftarrow$ \textbf{stl}(ts.transformed)
\ELSE
    \STATE seasonal $\leftarrow$ 0
    \STATE [trend, remainder] $\leftarrow$ \textbf{loess}(ts.transformed)
\ENDIF
\STATE ts.boot[1] $\leftarrow$ ts
\FOR{i $\leftarrow$ 2 to $Bs$}
    \STATE reminders.boot[i] $\leftarrow$ mbb.bootstrap(remainder, $l$)
    \STATE ts.boot.transformed[i] $\leftarrow$ trend + seasonal + reminders.boot[i]
    \STATE ts.boot[i] $\leftarrow$ \textbf{InvBoxCox}(ts.boot.transformed[i], $\lambda$)
\ENDFOR
\RETURN ts.boot
\end{algorithmic}
\end{algorithm}

\subsubsection{ETS Based Complex Seasonal Patterns}
Most canonical works do not consider and handle the complex seasonal patterns in the time series which are nested. Meanwhile, they are not capable of handling the non-linearity existing in the series. Besides, a large quantity of exponential smoothing models cannot cope with non-linearity problem and suffer from the overparameterization. So, \textbf{TBATS} \cite{de2011forecasting} is a innovations space modeling framework used to alleviate these problems, which stands for \textbf{T}rigonometric seasonality, \textbf{B}ox-Cox transformation, \textbf{A}RMA errors,  \textbf{T}rend and  \textbf{S}easonal components. Firstly, TBATS extend the Holt-winter addictive model with the Box-cox transformation, ARMA errors, and $T$ seasonal patterns formulated as following:
\begin{align}
    y_t^{(\lambda)} &= \ell_{t-1} + \phi b_{t-1} + \sum_{i=1}^Ts_{t-m_i}^{(i)} + d_t, \\
    \ell_t &= \ell_{t-1} + \phi b_{t-1} + \alpha d_t,\\
    b_t &= \phi b_{t-1} + \alpha d_t,\\
    d_t &= \sum_{i=1}^p \varphi_i d_{t-i} + \sum_{i=1}^q \theta_i \varepsilon_{t-i} + \varepsilon_t, \\
    s_t^{(i)} &= \sum_{j=1}^{k_i} s_{j,t}^{(i)} \\
    s_{j,t}^{(i)} &= s_{j,t-1}^{(i)} \cos (\omega_i) + s_{j,t-1}^{*(i)} \sin (\omega_i) + \gamma_1^{(i)}d_t \\
    s_{j,t}^{*(i)} &= -s_{j,t-1}^{(i)} \sin (\omega_i) + s_{j,t-1}^{*(i)} \cos (\omega_i) + \gamma_2^{(i)}d_t \\
    \omega_i &= \frac{2\pi j}{m_i}
\end{align}
where $y_t^{(\lambda)}$ is the time series after Box-cox transformation at time step $t$, $\{m_i\}_{i=1}^T$ are the seasonal periods, $T$ is the number of seasonalities, $b_t$ is the trend with damping parameter $\phi$, $\ell_t$ represents the local level in period $t$, $s_t^{(i)}$ denotes the $i$-th seasonal component at time $t$, $d_t$ is an ARMA($p,q$) process, and $\varepsilon_t$ is the Gaussian white noise process with zero-mean and constant variance. $\alpha, \beta, \gamma_1^{(i)}, \gamma_2^{(i)}$ are smoothing parameters. $k_i$ is the amount of harmonics for $i$-th seasonal period. We maxmimize the likelihood and AICc to find the best model.

For TBATS model, each seasonal pattern is modeled by a trigonometric representation based on Fourier series. Meanwhile, it only needs two seed states without involving the length of the period, and it also can model the non-integer seasonal lengths. To specify the model's structure, we denote model with its relevant arguments as TBATS$\left(\lambda, \{p,q\}, \phi, \{<m_1, k_1>, <m_2, k_2>, \dots, <m_T, k_T>\}\right)$


\subsection{Ensemble Learning}
\label{sec:ensemble}
For the standard supervised learning problem, the training examples tend to be involved with some random noise leading to the incorrect of the regressors or classifiers. To alleviate this problem, one of the practical techniques called ensemble learning \cite{dietterich2000ensemble} is adopted in many fields. An ensemble of regressors is a set of regressors whose individual predictions are combined in some way to predict the new points or series. The more diverse the regressors, the more accurate the forecast can be. It has been verified by Bates and Granger \cite{bates1969combination} that the combination forecasts often lead to better forecast accuracy, and use a simple average has proven hard to beat. Suppose we have a set of candidate models $\mathcal{M} = \{M_1, M_2, \cdots\}$, we can formulate this process as
\begin{align}
    \hat{y}_{t+h}^{\mathcal{M}} = \sum_{i=1}^{|\mathcal{M}|}p_{t+h}^{(M_i)}\hat{y}_{t+h}^{(M_i)},
\end{align}
where $\hat{y}_{t+h}^{\mathcal{M}}$ is the final prediction, $p_{t+h}^{\mathcal{M}}$ is the weight of model $M_i$ at time step $t+h$, and $\hat{y}_{t+h}^{(M_i)}$ is the prediction generated by the model $M_i$ at time step $t+h$. $|\mathcal{M}|$ is the number of models in the set $\mathcal{M}$. In order to estimate the weight $p_{t+h}^{\mathcal{M}}$, we would use Online Gradient Descent (OGD) method \cite{zinkevich2003online} to calculate the weight at each time step, and we name this model as OGD for abbreviation.

\section{Experiments}
\label{sec:experiments}
\subsection{Evaluation Metrics}
For evaluation, we use the the Mean Average Precision Error (sMAPE) to measure the prediction errors due to the advantage of being unit-free. Formally, sMAPE is formulated as
\begin{equation*}
    \text{MAPE} = \frac{100}{T}\sum_{t = 1}^T \frac{|y_t - \hat{y}_t|}{y_t}.
\end{equation*}
where $y_t$ is the actual value of time series at time $t$, and $\hat{y}_t$ is the predicted value at time $t$.  Besides, the symmetric Mean Absolute Percentage Error (sMAPE) can also evaluate models' performance and the formula is
\begin{equation*}
    \text{sMAPE} = \frac{200}{T}\sum_{t = 1}^T\frac{|y_t - \hat{y}_t|}{|y_t| + |\hat{y}_t|},
\end{equation*}
In the same time, the Root Mean Square Error (RMSE) is adopted to measure the ability of model prediction. Besides, we evaluated our results using Mean Percentage Error (MPE) which can be written as
\begin{equation*}
    \text{MPE} = \frac{100}{T}\sum_{t=1}^T\frac{y_t - \hat{y}_t}{y_t}.
\end{equation*}

\subsection{Evaluation Settings}
For traffic accidents time series, we divide the series into three consecutive time periods: 80\% time of time series for training, last 5\% time of time series for testing, and the rest 15\% are used for validation.

\begin{table}[]
\centering
\begin{tabular}{l|c|c|c|c}
\hline\hline
\textbf{Model}               & \textbf{RMSE}    & \textbf{MPE}   & \textbf{MAPE}  & \textbf{sMAPE} \\ \hline
ARI(2,1)                     & 199.667          & 3.077          & 3.801          & 3.921          \\ \hline
IMA(1,1)                   & 153.122          & -0.077         & 3.429          & 3.435          \\ \hline
ARIMA(1,1,1)                 & 193.023          & 2.751          & 3.686          & 3.804          \\ \hline
Simple Exponential Smoothing & 193.816          & 2.824          & 2.735          & 3.844          \\ \hline
ETS(A, N, A)                 & 124.676          & 2.096          & 2.728          & 2.771          \\ \hline
Bagged ETS                   & 133.530          & 2.341          & 2.966          & 3.017          \\ \hline
Dynamic Harmonic Regression ($K=5$) & 121.624          & 1.974          & 2.661          & 2.701          \\ \hline
TBATS(1,\{0,0\},-,\{<12,5>\})        & 120.149          & 1.735          & 2.561          & 2.593          \\ \hline
OGD              & \textbf{114.143} & \textbf{1.439} & \textbf{2.293} & \textbf{2.322} \\ \hline
\end{tabular}
\vspace{0.05in}
\caption{Performance comparisons with different models on validation series.}
\label{tab:results}
\end{table}

\subsection{Quantitative Results}
We compare fitting results among the proposed models, and test these models on the validation series. The results are reported in Table \ref{tab:results}. Several interesting observation can be found as follows.
\begin{itemize}
    \item IMA(1,1) model outperforms most of the competing ARIMA models by a large margin in the validation set. IMA(1,1) is better than the rest at least 30 on RMSE. This is because if we look at the forecast series generated by the IMA(1,1), we can see a clear increasing trend that is more compatible with the actual trend. However, for ARI(2,1) and ARIMA(1,1,1), these two models predict the series which perform constantly.
    \item In the column of measuring MPE, only IMA(1,1) achieves negative value which closes to zero. It is on the grounds that the true series fluctuates around the IMA(1,1) predicted series. Therefore, according to the formula of MPE, we can see it can result in the negative value in the numerator if $y_t < \hat{y}_t$ happens. So the smaller MPE cannot indicate the model fits better than other models.
    \item The ETS(A, N, A) achieves better performance in all measurements than the Bagged version of ETS due to the random sampling noise in the generated bootstrapped series. As a consequence, ETS would result in more accurate predictions in the forecast.
    \item Dynamic harmonic regression model and TBATS(1,\{0,0\},-,\{<12,5>\}) model obtain larger gains on each evaluation metrics. The shared characteristic of these two models is that they both utilize the Fourier terms to model the complex seasonality and non-linearity. For the estimated number of periods, it is surprising that both models give $K=5$ as the number of periods, which verifies that the original series involved with complex seasonality and non-linearity.
    \item For the final model with ensemble method, we select models with small RMSE for the base model for final OGD model which consists of IMA(1,1), ETS(A, N, A), DHR ($K=5$), and TBATS(1,\{0,0\},-,\{<12,5>\}).
    Meanwhile, the best performance strongly supports our intuition that different models would alleviate the random noise in the training series leading to a more accurate forecast.
\end{itemize}{}

\begin{figure}
     \centering
     \includegraphics[width=\textwidth]{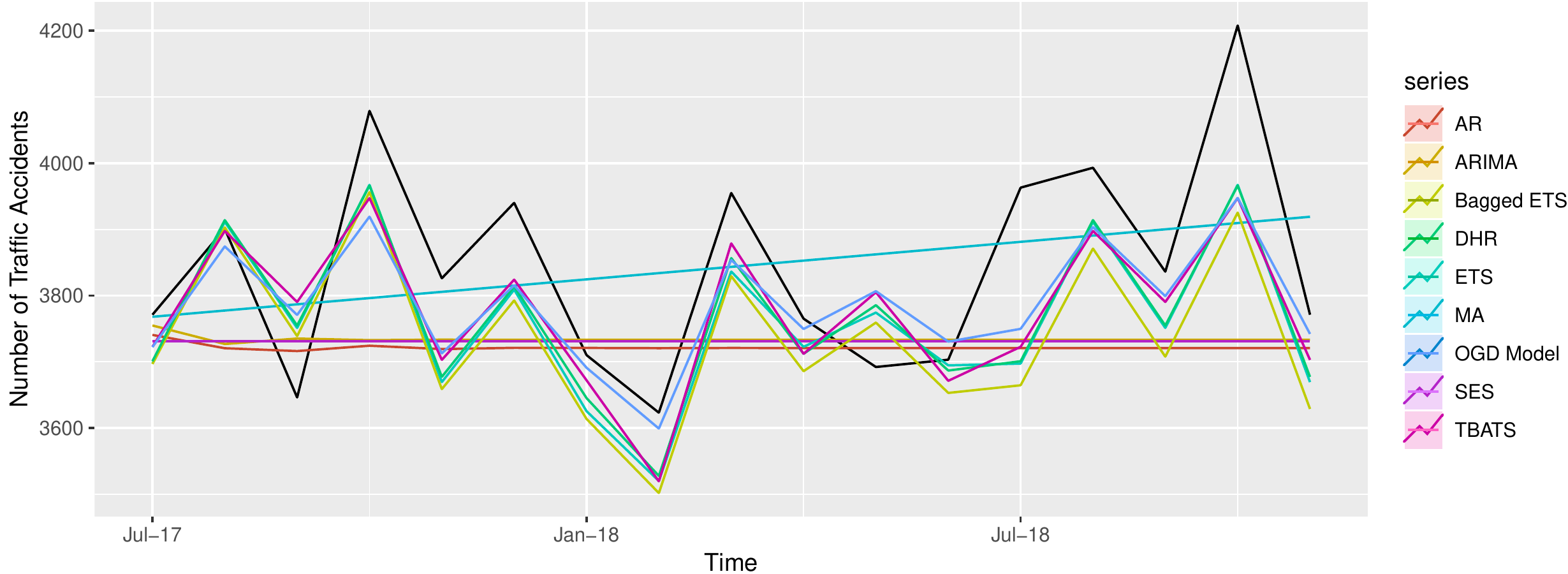}
     \caption{Visualization of different models' predictions on the validation series.}
    \label{fig:result_visualize}
\end{figure}

Besides, we visualize the forecast series and the ground truth series in Figure \ref{fig:result_visualize}. For most predictions, their trends are similar to ground truth series, but these predictions tend to underestimate the actual number of accidents. It is possible that prediction errors would result from other factors such as policy, migration, and weather condition, etc.

\section{Model Diagnostics}
\label{sec:diagnostic}

In this section, we would examine some good models mentioned in Table \ref{tab:results} to verify the results and choose the ultimate model for forecasting.

\subsection{Residual Analysis of IMA(1,1) Model}
\label{sec:ima_check}
\begin{figure}
     \centering
     \includegraphics[width=\textwidth]{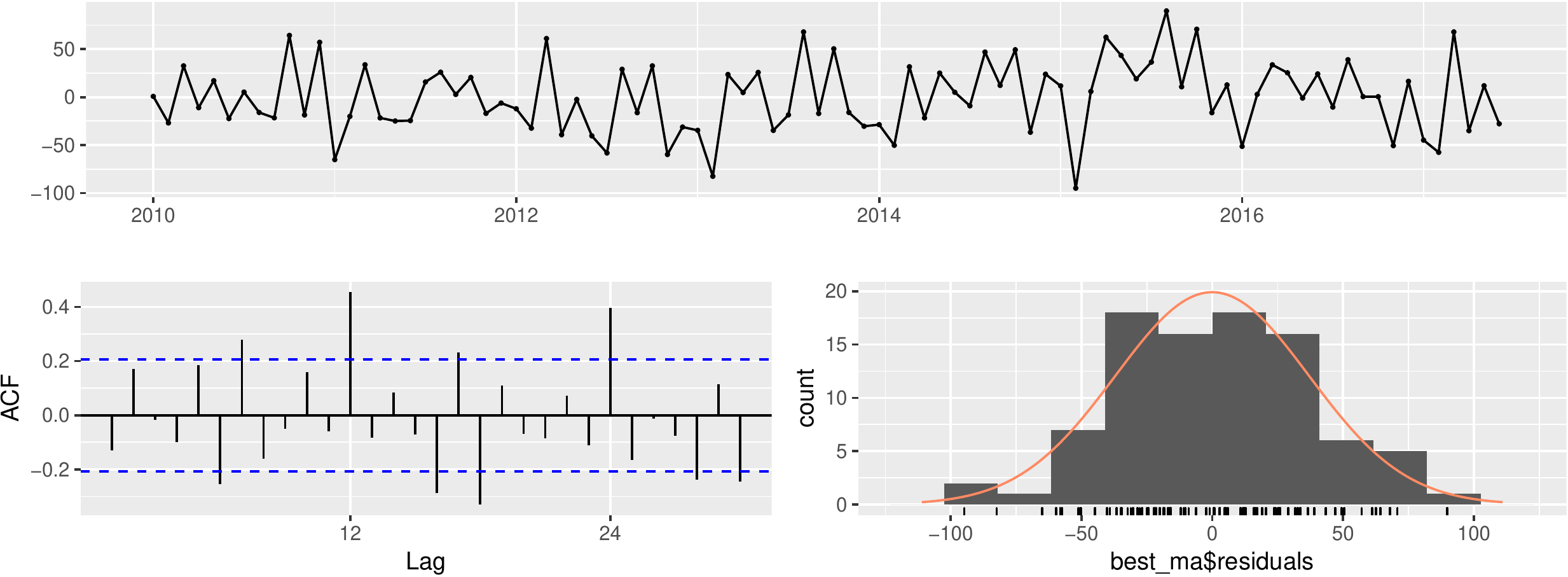}
     \caption{Normality and independence assumptions checking for IMA(1,1) model.}
    \label{fig:residual_ima}
\end{figure}

Since we start from the ARIMA family model, we observe that IMA model is more proper than ARI or ARIMA model due to the fact that the first-order differenced series is stationary. So, we fit IMA(1,1) model represented as following:
\begin{align}
    \hat{y}_t' = 1.9149 -0.7257 y_{t-1}',
\end{align}{}
where $y_{t}'$ is the first-order differenced series. To assess the fitted IMA(1,1) model visually, we draw the figure of ACF plot of residuals generated by fitted IMA(1,1) model in Figure \ref{fig:residual_ima}. From the ACF plot, we can clearly see that there exist some sharp spikes at lag 6, 7, 12, 16, 18, and 24, which indicates that the residuals are still correlated with each other. To verify our observation, Ljung-Box test shows that p-value is $1.318\times 10^{-10}$ which is really close to 0. Therefore, we should reject the null hypothesis that the residuals are uncorrelated. We have sufficient evidence to conclude that the residuals are correlated. In consequence, the series cannot be well explained by IMA(1,1) model.

\subsection{Residual Analysis of ETS(A, N, A) Model}

\begin{figure}
     \centering
     \includegraphics[width=\textwidth]{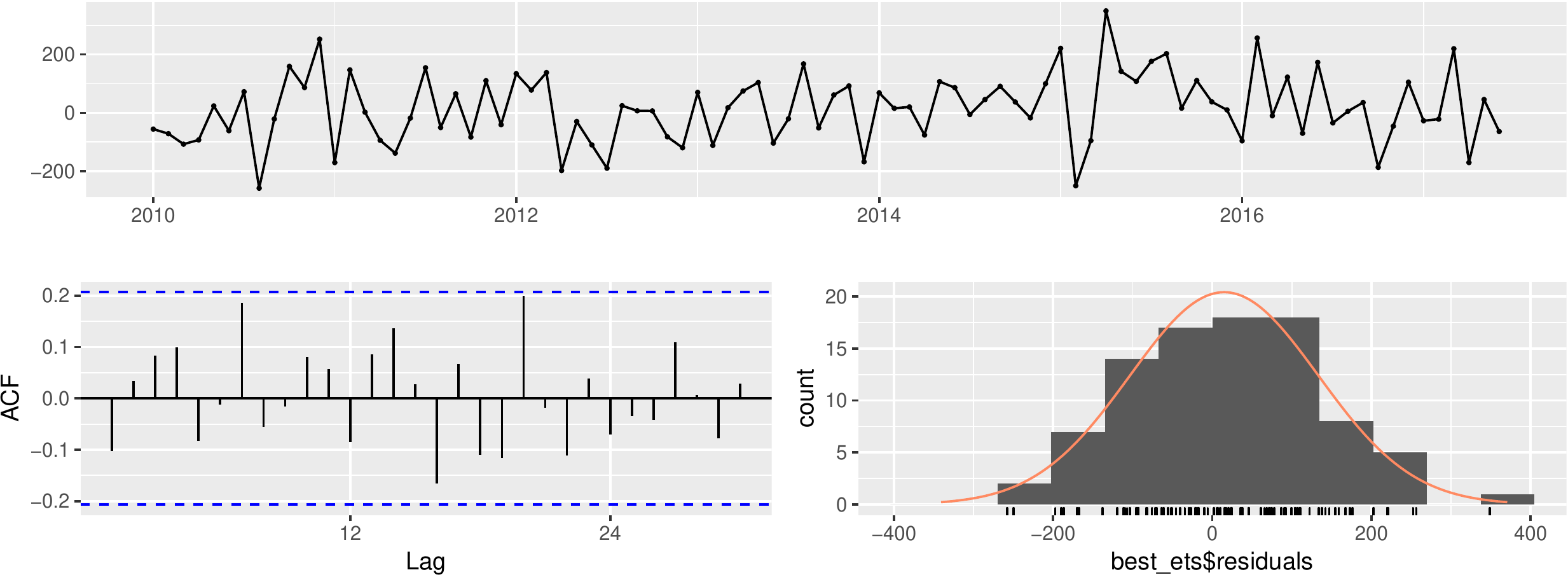}
     \caption{Normality and independence assumptions checking for ETS(A, N, A).}
    \label{fig:residual_ets}
\end{figure}

From the fitted model, it can be pointed out that the ETS(A, N, A) model is equivalent to ARIMA$(0, 1, 1)(0, 1, 0)_1$. It shows that it not only considers the IMA(1, 1) process analyzed in Section \ref{sec:ima_check} but also takes the seasonal components into account. Consequently, when we check the residuals of fitted ETS(A, N, A) model, we have enough evidence to conclude that the residuals are normally distributed due to the p-value 0.9916. Simultaneously, the ACF plot in Figure \ref{fig:residual_ets} demonstrates that all the autocorrelations are within the error bounds showing uncorrelation characteristic. However, to quantitatively uncorrelation, we use Ljung-Box test to check whether the residuals are uncorrelated from each other. Ljung-Box test generates the p-value 0.00214, showing that we should reject the null hypothesis that the data is uncorrelated under the significance level 0.05. Therefore, the model can not explain the series well.

\subsection{Residual Analysis of DHR ($K=5$) Model}

\begin{figure}
     \centering
     \includegraphics[width=\textwidth]{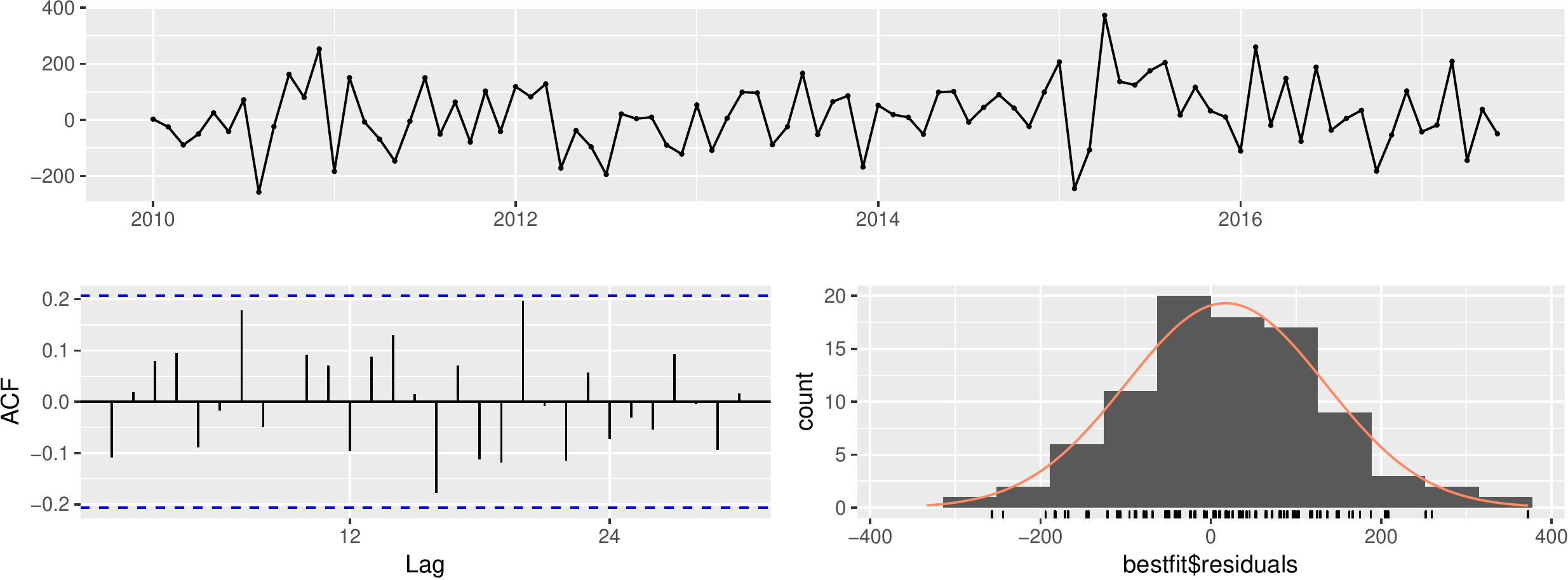}
     \caption{Normality and independence assumptions checking for DHR model.}
    \label{fig:residual_dhr}
\end{figure}

With Fourier items being added, the ARIMA model will then have deterministic, but smoothed, seasonal effects. We choose the best $K$ according to AICc criteria. So after taking the Fourier item into consideration, we can see all the values in ACF plot are within the error bounds, which indicates that the residuals are uncorrelated. Meanwhile, we use Shapiro-Wilk test whether the residuals are normally distributed or not. It results in the p-value 0.9618, which points out the residuals are normally distributed. Moreover, we conduct the runs test to check the independence assumption which results in p-value 0.193 showing the residuals are independent. However, when we verify the autocorrelation with Ljung-Box test, we achieve the small p-value 0.007595 which means the residuals are still correlated indicating this model may not be a good fit.

\subsection{Residual Analysis of TBATS(1,\{0,0\},-,\{<12,5>\}) Model}

\begin{figure}
     \centering
     \includegraphics[width=\textwidth]{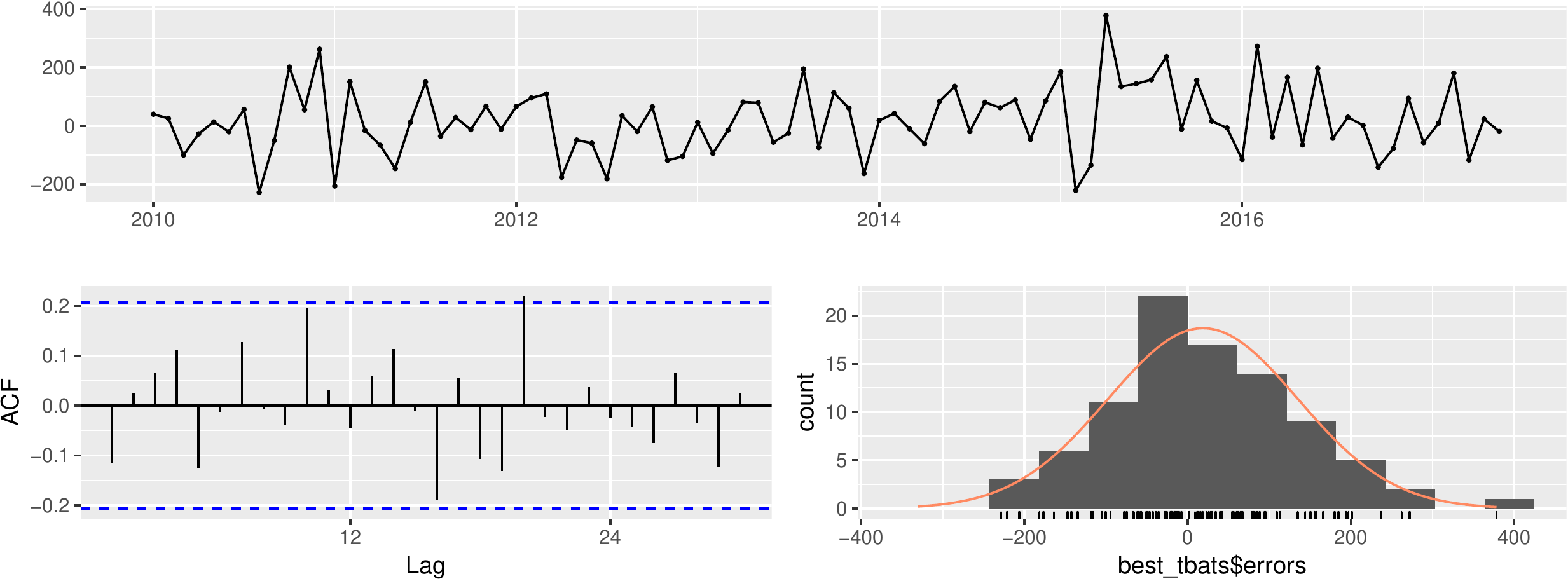}
     \caption{Normality and independence assumptions checking for TBATS(1,\{0,0\},-,\{<12,5>\}) model.}
    \label{fig:residual_tbats}
\end{figure}
TBATS models the time series data from exponential smoothing method which is different from ARIMA family model due to the idea of decomposition. However, the shared feature of TBATS and Dynamic Harmonic Regression is that both methods utilize the Fourier items to model the complex seasonality. To check the validity of the model, we plot the residuals and their ACF in Figure \ref{fig:residual_tbats}. Similarly, following the protocol of checking model, we get the p-value 0.7057 from Shapiro-Wilk test, showing the residuals can be regarded as normally distributed. Besides, we conduct the run test for checking the independence which results in p-value 0.0382 showing that we do not have sufficient evidence to conclude that residuals are independent. Thus, the independence assumption is not satisfied. So the model cannot well model the series.

\subsection{Composition Analysis and Residual Analysis of OGD Model}
\begin{figure}
     \centering
     \includegraphics[width=\textwidth]{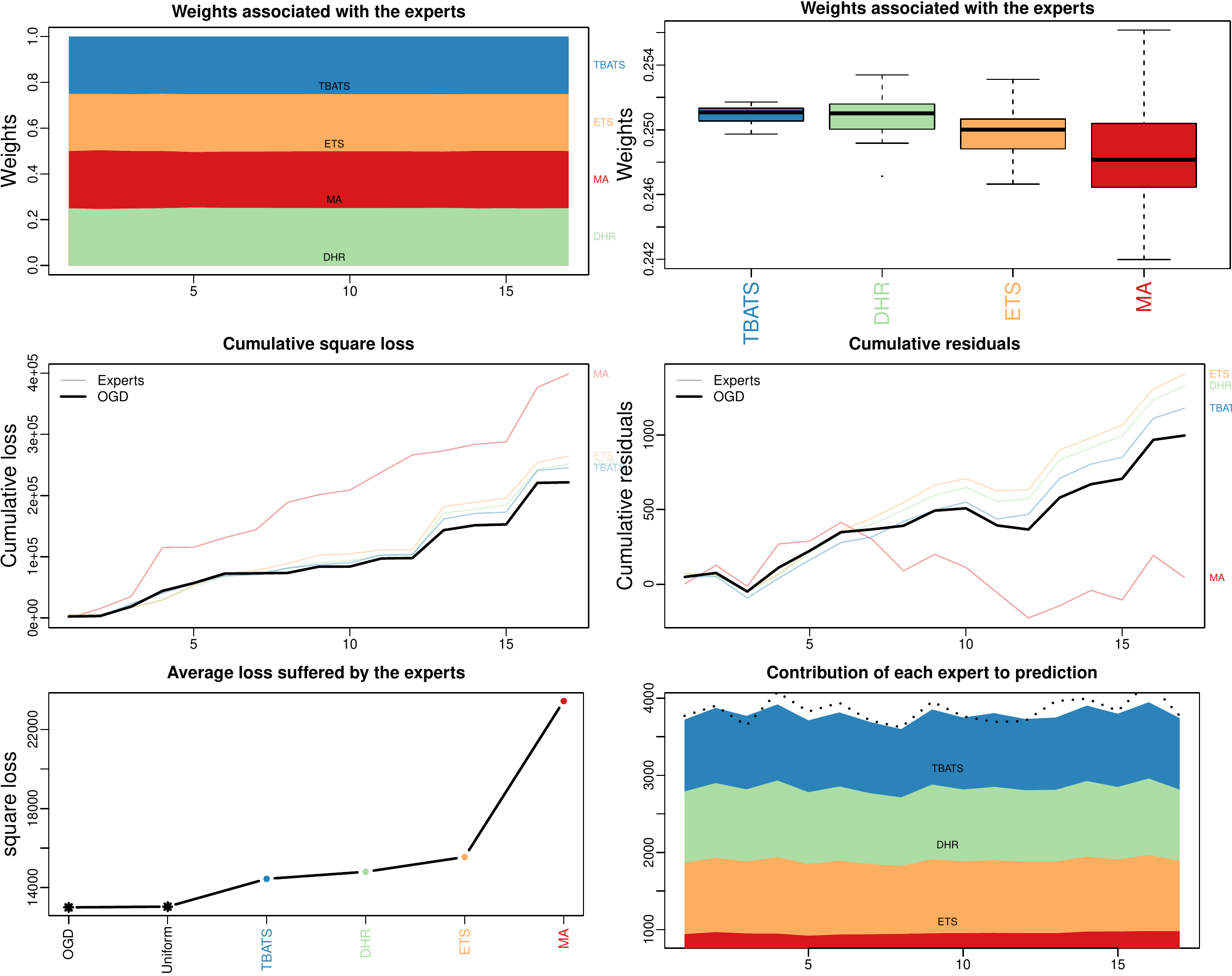}
     \caption{Visualization of the composition of OGD model by combining IMA(1,1), ETS(A, N, A), DHR ($K=5$), and TBATS(1,\{0,0\},-,\{<12,5>\}).}
    \label{fig:ogd_weight}
\end{figure}

Through ensemble learning method, we build OGD model using IMA, ETS, DHR, and TBATS. Before we consider the residual of the fitted model, we visualize the weights of each model in OGD model as Figure \ref{fig:ogd_weight} represents. In the subfigure on the left upper corner, we can clearly see that the total weights distribution is like uniform distribution. However, there are slight differences between uniform distribution since if we glance at the subfigure in the left lower corner since it shows that OGD method is slightly better than uniform method. At the same time, the ensemble method enables random errors to decrease as much as possible.

\begin{figure}
     \centering
     \includegraphics[width=\textwidth]{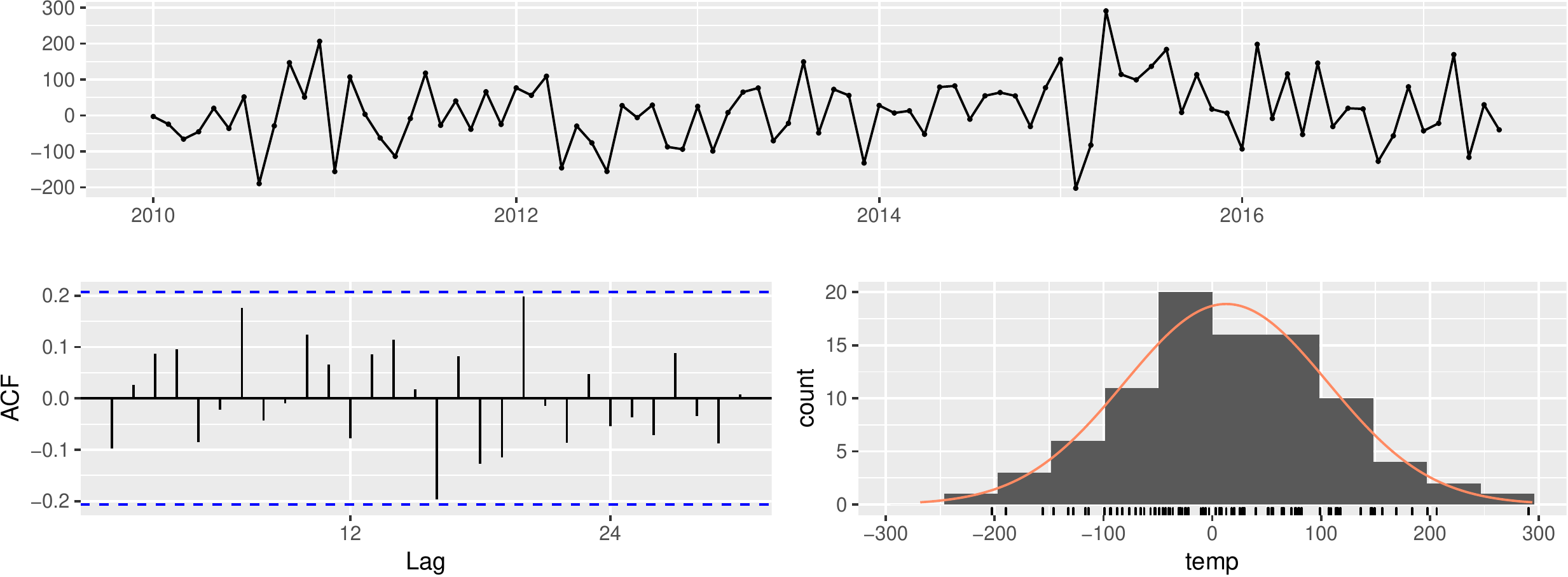}
     \vspace{0.025in}
     \caption{Normality and independence assumptions checking for OGD model.}
    \label{fig:residual_ogd}
\end{figure}

Furthermore, we analysis the residual of fitted model for the sake of verifying its effectiveness. According to Figure \ref{fig:residual_ogd}, we can observe the ACF plot that there is no spike in the plot and all the autocorrelation values are within the error bounds, which means that residuals of fitted OGD model are independent. To quantitatively verify the uncorrelation, we conduct Ljung-Box test to check whether the residuals are uncorrelated from each other. It shows that test statistic $\chi^2 = 24.673$ with degree of freedom 20 has the p-value $=0.2142$ indicating that we failed to reject the null hypothesis, which proves the uncorrelation. Apart from checking the uncorrelation, we also verify the normality of the residuals. In order to show the normality, we use Shapiro-Wilk test. It comes that the p-value is 0.9784 and means that the residuals are normally distributed. So we can regard these residuals as the errors generated by white noise process. Therefore, OGD models the whole time series quite well compared to other methods.

\section{Forecasting}
\label{sec:forecast}
In this section, we forecast the number of traffic accidents in the future using the OGD model mentioned in Section \ref{sec:ensemble}. For better understanding the accuracy of prediction, we predict the series on test series which dates from December 2018 to May 2019. In addition, we train our model on training series and validation series  The testing results are summarized in Table \ref{tab:forecast}. 

\begin{table}[]
\begin{center}
\resizebox{1.0\linewidth}{!}{%
\begin{tabular}{l|c|c|c|c|c|c|c|c|c|c}
\hline\hline
                                   & Dec. 2018 & Jan. 2019 & Feb. 2019 & Mar. 2019 & Apr. 2019 & May 2019 & RMSE                     & MPE                     & MPAE                   & sMPAE                  \\ \hline
Ground Truth                       & 3699      & 3602      & 3597      & 3966      & 3766      & 4022     & \multirow{4}{*}{176.573} & \multirow{4}{*}{-3.573} & \multirow{4}{*}{4.256} & \multirow{4}{*}{4.151} \\ \cline{1-7}
Prediction (Mean)                  & 3936.8    & 3870.3    & 3788.6    & 4020.6    & 3875.4    & 3875.4   &                          &                         &                        &                        \\ \cline{1-7}
SE ($\hat{y}_t$)$_{\alpha = 0.80}$ & 232.7     & 234.4     & 253.8     & 261.4     & 268.9     & 276.1    &                          &                         &                        &                        \\ \cline{1-7}
SE ($\hat{y}_t$)$_{\alpha = 0.95}$ & 357.3     & 359.9     & 289.7     & 401.5     & 412.9     & 424.1    &                          &                         &                        &                        \\ \hline
\end{tabular}
}
\end{center}
\caption{Forecast results on the testing series with OGD Model compared to the ground truth series. The results include mean prediction value and standard error under 80\% and 95\% confident level.}
\label{tab:forecast}
\end{table}

As Table \ref{tab:forecast} presents, the predict values tend to be slightly higher than the actually observed values about 200, this might be caused by the population and economics in the Log Angeles, which suggests our model can explain adequately the traffic accident data. On the other hand, we can see as time pass by, the range of confidence intervals tends to be larger which means more uncertainty in future prediction, and actual observation values are in the interval showing the correctness of our method.

\section{Discussion}
\subsection{Summary}
In the paper, we tried to analyze the pattern of traffic accident happened last few years in Los Angeles. We investigated the time series we are interested in terms of statistical characteristics such as trend, seasonality, correlation, and skewness, etc. We also selected a set of candidate models and proposed ensemble method. Then, we fitted the models on the training series and validated these models on validation series to select the best model. Meanwhile, residual diagnostics of different models were fully presented. The ensembled model OGD showed the best fit among candidate models used for final forecasting on the test series.

\subsection{Future Work}
For further modeling the traffic accident time series, we can take the characteristic of the data. In other words, the distribution of traffic accident follows a Poisson distribution or negative binomial (NB) distribution. We can utilize these assumptions to adopt the Poisson regression or negative binomial regression model to fit the data. Moreover, the combination of  generalized linear model and a stationary model can be more effective in modeling the data.


\bibliographystyle{plainnat}
\bibliography{reference}

\newpage
\appendix

\afterpage{%
\newgeometry{left=1cm,right=1cm,top=3cm,bottom=3cm}
\begin{figure}[t!]
     \centering
     \begin{subfigure}[t]{.5\textwidth}
        \centering
        \includegraphics[width=0.9\textwidth]{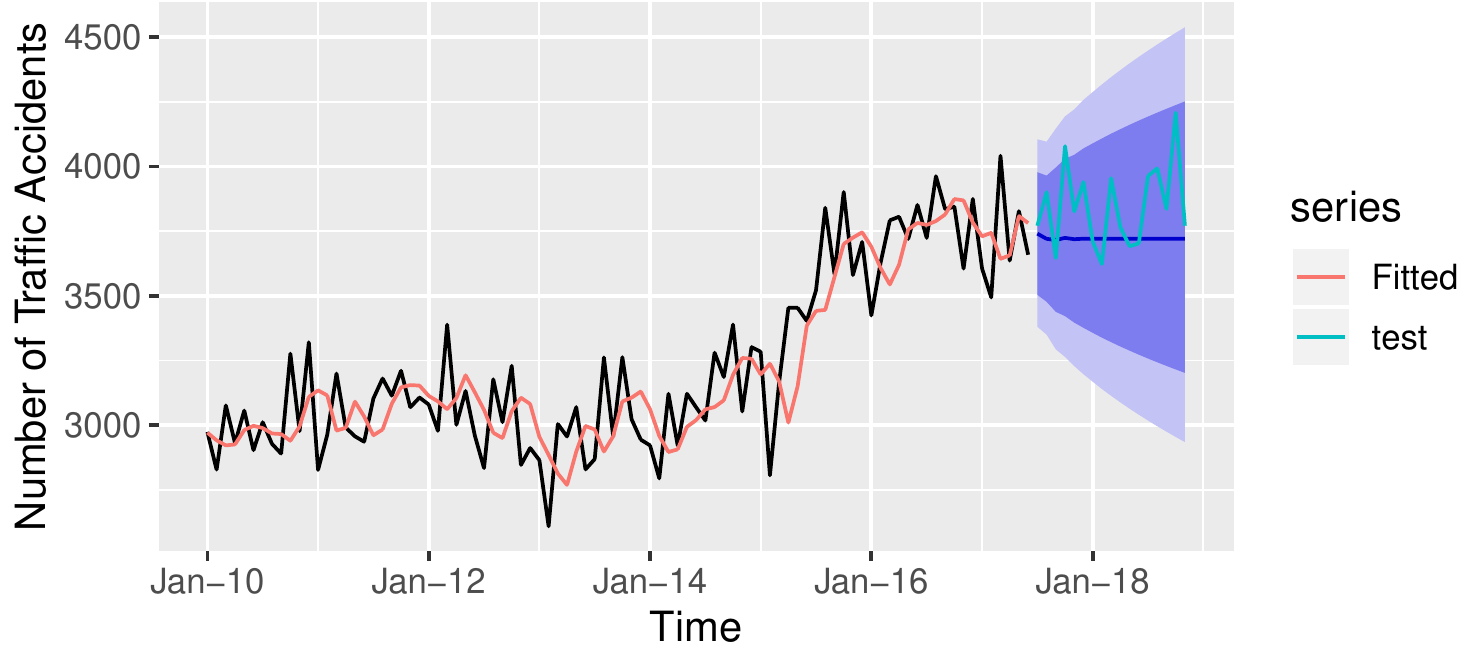}
        \caption{ARI(2, 1)}
    \end{subfigure}%
    ~
    \begin{subfigure}[t]{.45\textwidth}
        \centering
        \includegraphics[width=\textwidth]{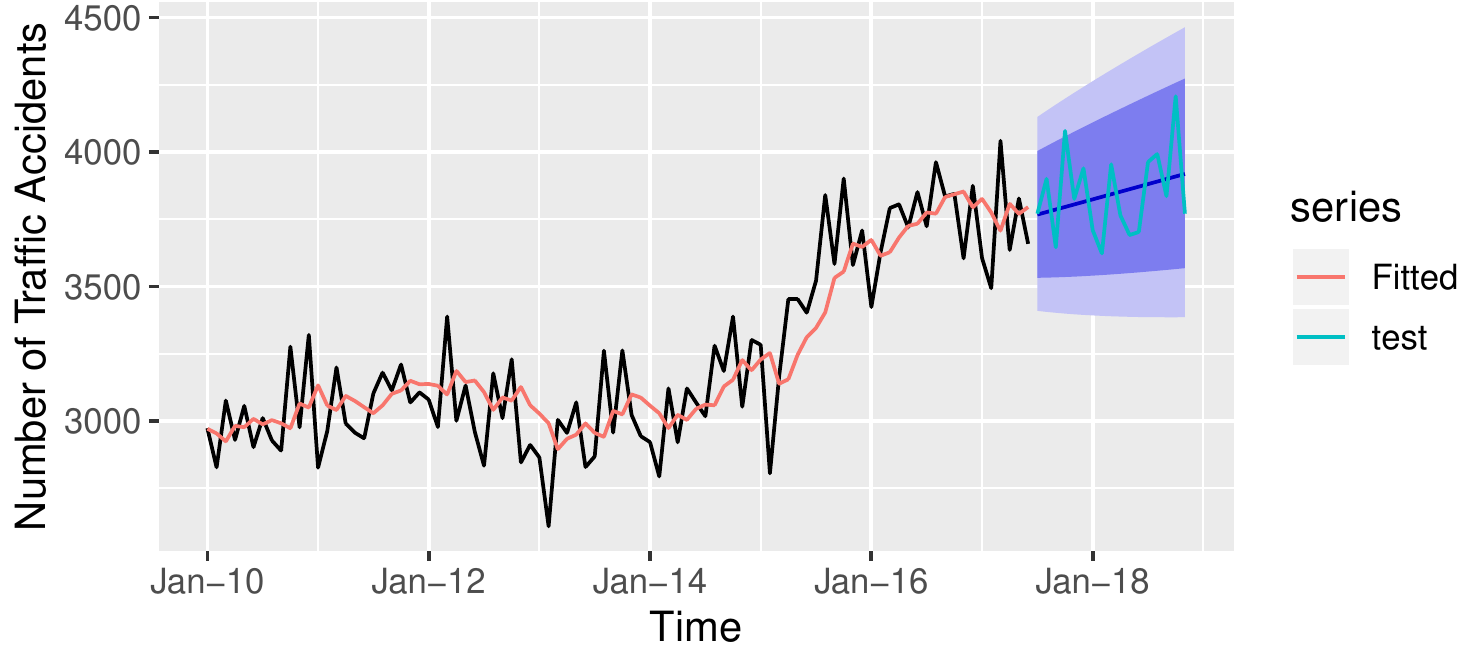}
        \caption{IMA(1, 1)}
    \end{subfigure}
    \vspace{0.5cm}
    
    \begin{subfigure}[t]{.5\textwidth}
        \centering
        \includegraphics[width=0.9\textwidth]{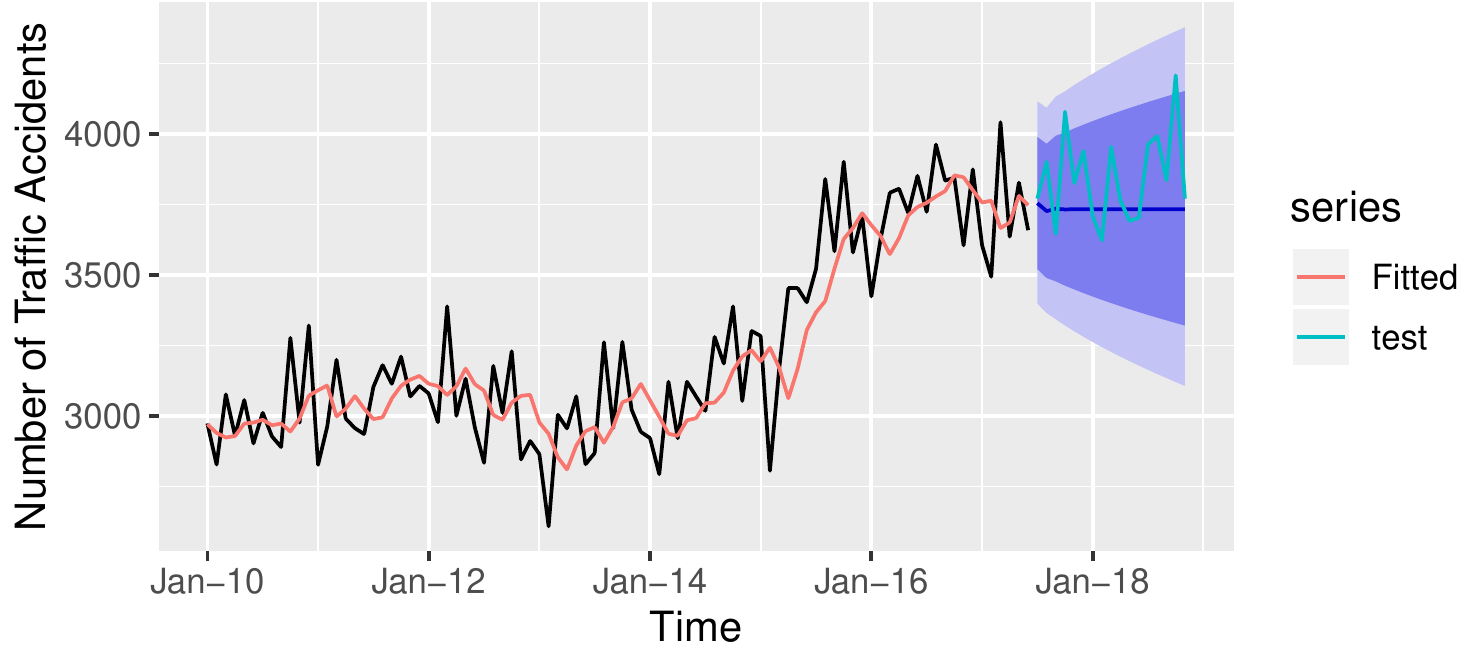}
        \caption{ARIMA(1, 1, 1)}
    \end{subfigure}
    ~
    \begin{subfigure}[t]{.45\textwidth}
        \centering
        \includegraphics[width=\textwidth]{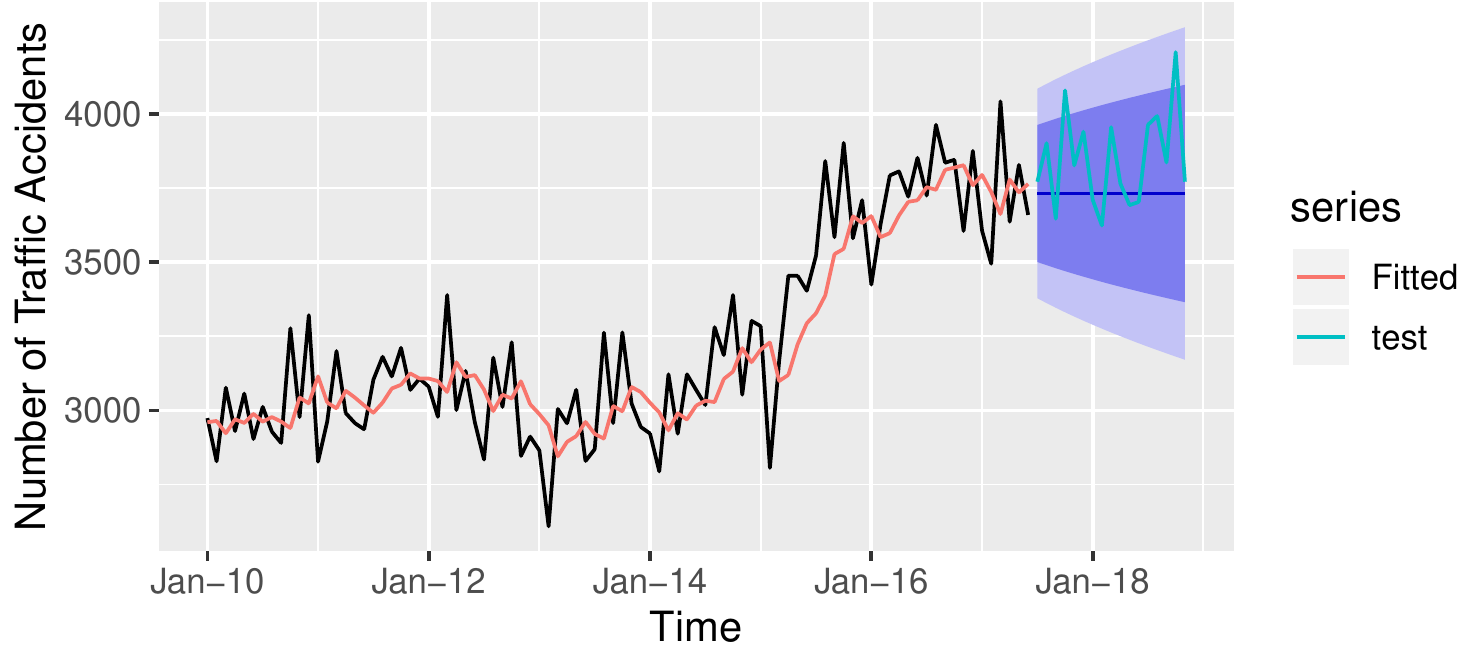}
        \caption{Simple Exponential Smoothing}
    \end{subfigure}
    \vspace{0.5cm}
    
    \begin{subfigure}[t]{.5\textwidth}
        \centering
        \includegraphics[width=0.9\textwidth]{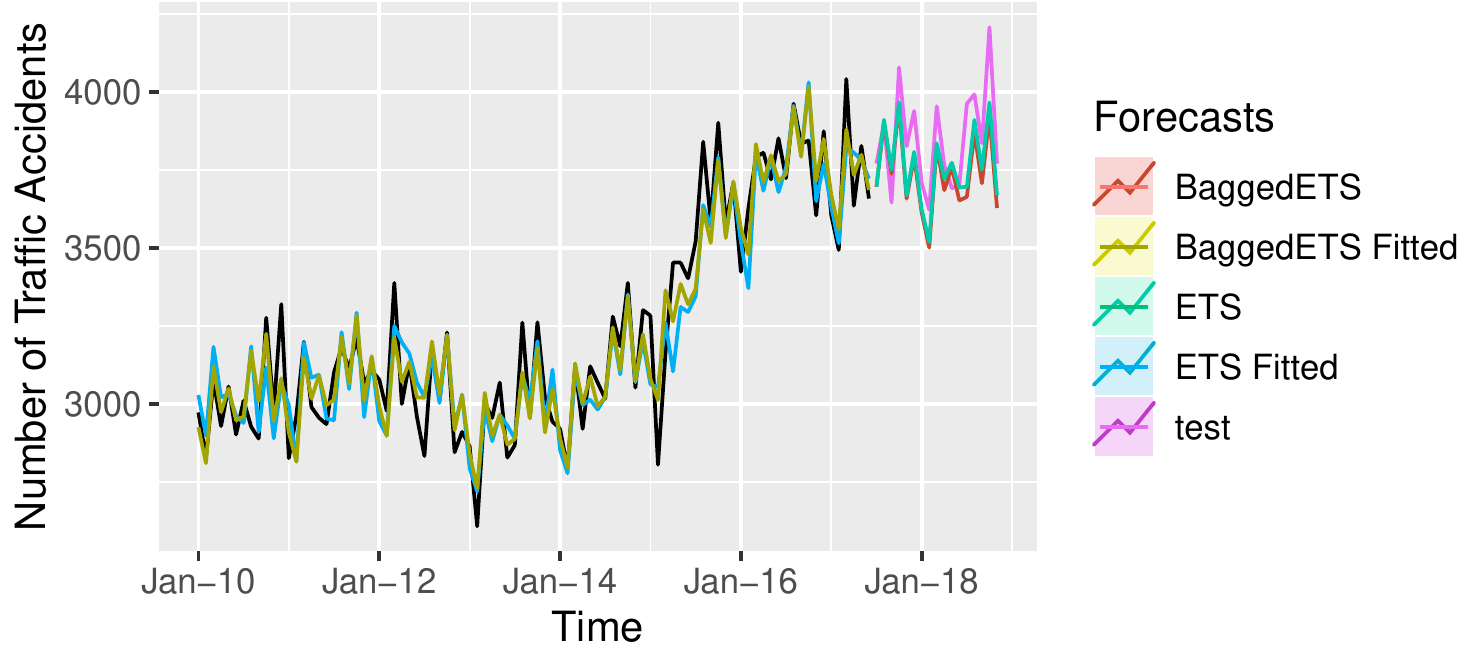}
        \caption{ETS and Bagged ETS}
    \end{subfigure}
    ~
    \begin{subfigure}[t]{.45\textwidth}
        \centering
        \includegraphics[width=\textwidth]{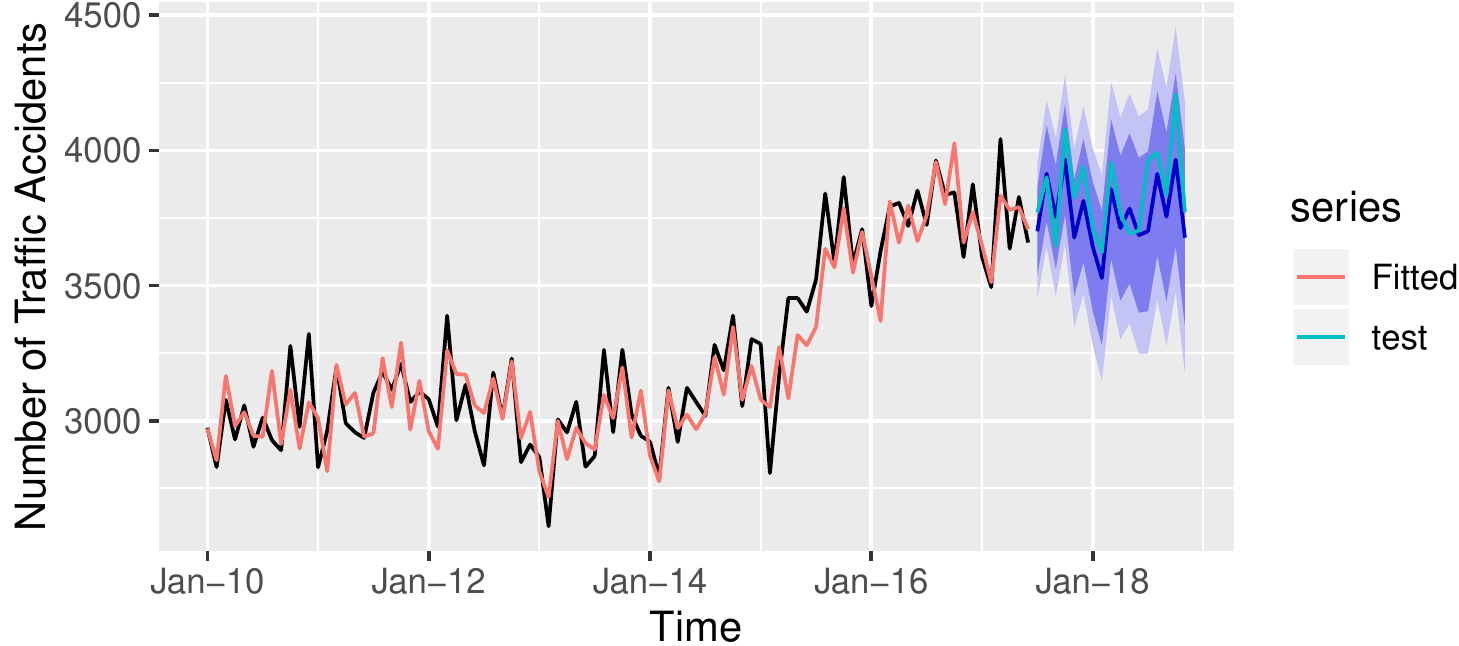}
        \caption{Dynamic Harmonic Regression ($K=5$)}
    \end{subfigure}
    \vspace{0.5cm}

    \begin{subfigure}[t]{.5\textwidth}
        \centering
        \includegraphics[width=0.9\textwidth]{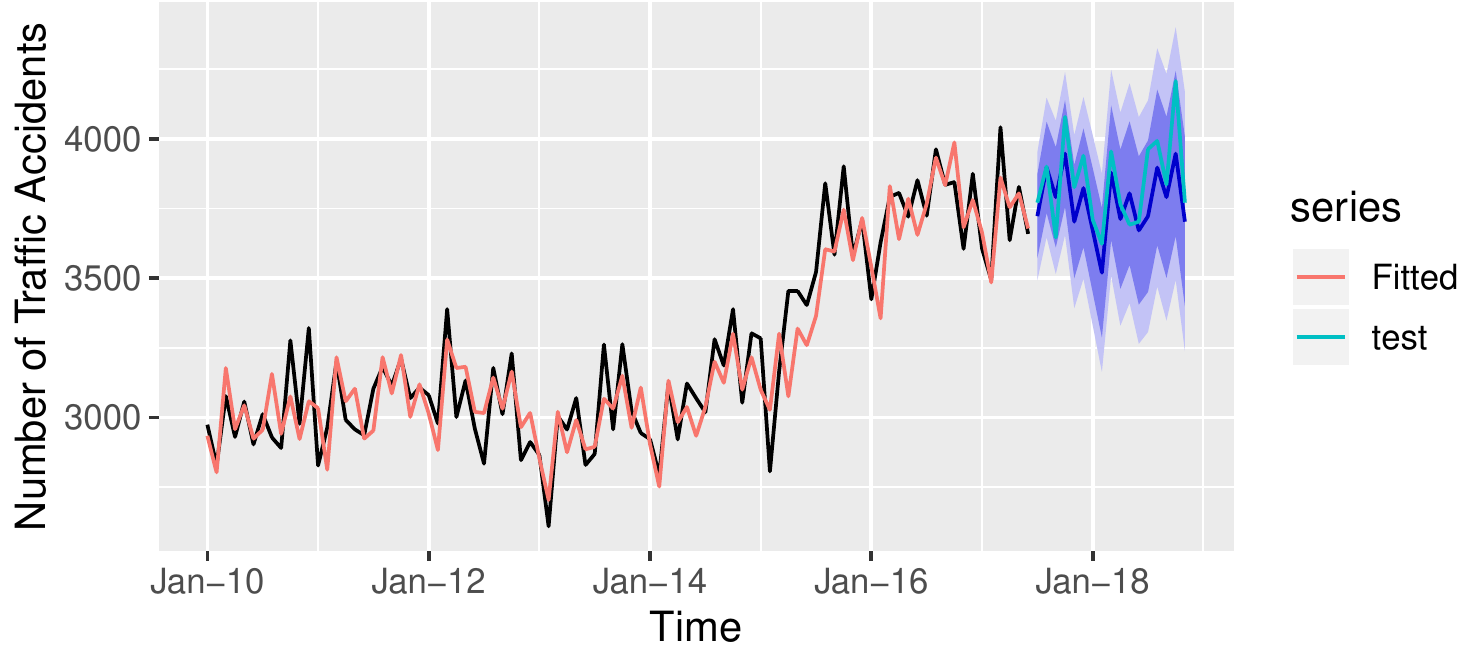}
        \caption{TBATS(1,\{0,0\},-,\{<12,5>\})}
    \end{subfigure}
    ~
    \begin{subfigure}[t]{.45\textwidth}
        \centering
        \includegraphics[width=\textwidth]{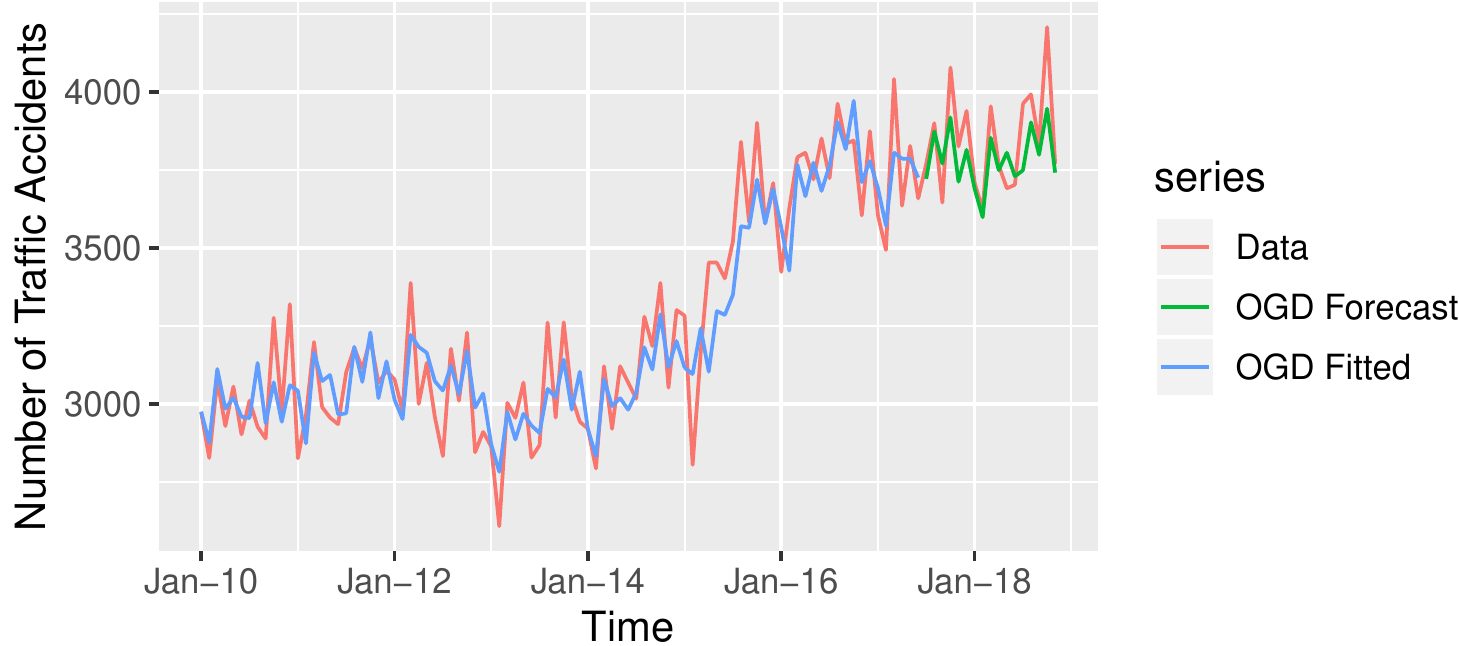}
        \caption{OGD Model}
    \end{subfigure}    
    \vspace{0.5cm}
    
    \caption{Candidate models performance on training and validation series.}
\end{figure}

\clearpage
\restoregeometry
}

\end{document}